\documentclass[11pt]{article}

\usepackage{amsmath,amsthm,amssymb,amsfonts,epsfig,xcolor,cite}

\newcommand{\be}{\begin{equation}}
\newcommand{\ee}{\end{equation}}
\newcommand{\ba}{\begin{eqnarray}}
\newcommand{\ea}{\end{eqnarray}}

\title{{{\sf Relational Lorentzian Asymptotically Safe Quantum Gravity:}\\
{\sf Showcase model}}
\author{
{\sf R. Ferrero}$^1$\thanks{{\sf
renata.ferrero@gravity.fau.de}},
{\sf T. Thiemann}$^1$\thanks{{\sf
thomas.thiemann@gravity.fau.de}}\\
}
\\
{\sf \normalsize  $^1$ Inst. for Quantum Gravity, FAU Erlangen -- N\"urnberg,}\\
{\sf \normalsize Staudtstr. 7, 91058 Erlangen, Germany}\\
}
\date{{\small\sf \today}}

\makeatletter
\@addtoreset{equation}{section}
\makeatother

\begin{document}

\maketitle

{\sf

\begin{abstract}
In a recent contribution we identified possible points of contact between the 
asymptotically safe and canonical approach to quantum gravity. The 
idea is to start from the reduced phase space (often called relational)
formulation of canonical quantum gravity which provides a reduced 
(or physical) Hamiltonian
for the true (observable) degrees of freedom. The resulting reduced phase 
space is then canonically quantised and one can construct the 
generating functional of time ordered Wightman (i.e. Feynman) 
or Schwinger distributions respectively from the corresponding
time translation unitary group or contraction semigroup respectively  
as a path integral. For the unitary choice 
that path integral can be rewritten in terms 
of the Lorentzian Einstein Hilbert action plus observable 
matter action and a ghost action.
The ghost action depends on the Hilbert space 
representation chosen for the canonical quantisation and a reduction 
term that encodes the reduction of the full phase space to the phase space 
of observavbles. This path integral can then be treated with the methods 
of asymptically safe quantum gravity in its {\it Lorentzian} version.

We also exemplified the procedure using a concrete,  minimalistic 
example namely 
Einstein-Klein-Gordon theory with as many neutral and massless scalar
fields as there are spacetime dimensions. However, no explicit calculations 
were performed. In this paper we fill in the missing steps. Particular
care is needed due to the necessary switch to Lorentzian signature  
which has strong impact on the convergence of ``heat'' kernel 
time integrals in the heat kernel expansion of the trace 
involved in the Wetterich equation and which requires different 
cut-off functions than in the Euclidian version. As usual we truncate 
at relatively low order and derive and solve the resulting flow equations 
in that approximation.  
\end{abstract}

\section{Introduction}
\label{s1}

The asymptotically safe quantum gravity (ASQG) \cite{1, 1.1, 1.2}
and canonical quantum gravity \cite{2, 2.1, 2.2, 2.3} programmes are both non-perturbative 
approaches with the common goal to synthesise Quantum Field Theory (QFT) 
and General Relativity (GR). However, there appear to be profound differences 
between the two at a very deep level:\\
1. While ASQG uses mostly Euclidian 
signature, CQG uses exclusively Lorentzian signature.\\ 
2. While ASQG employs background 
dependent methods, CQG is manifestly background independent.\\
3. While 
ASQG relies on truncations of the exact renormalisation 
flow (Wetterich) equations, no 
truncations are performed in CQG.\\ 
These differences are so drastic 
that very little contact between the two programmes has been established 
so far.

In a recent contribution \cite{3}
we have advertised the point of view that these differences are possibly 
not as unsurmountable as they appear to be. First of all, there are 
also Lorentzian versions of the Wetterich equation which were applied 
to matter quantum fields and gravity \cite{4, 4.1, 4.2, 4.3, 41, 41.1, 41.2,42,43, 44}. Next, the apparent background dependence 
of ASQG is mostly a misunderstanding if one uses the background techniques 
of ASQG properly. Namely, the so called background field method 
\cite{5} was invented   
for QFT without gravity as a tool to compute the effective action (i.e. 
the generating functional of connected, 1-PI, time ordered distributions)
which by itself is of course background independent and which results 
from the background dependent object by equating the background field 
with the current on which the background independent effective action 
depends. In order that this works, one must use unspecified background 
fields which are not subject to any particular restrictions such as 
symmetries. Finally, the truncations performed in ASQG are not 
required in principle but in practice in the sense of an approximation
scheme which, to the best of our 
knowledge, is the case for all known renormalisation procedures. In view 
of the fact that also in CQG renomalisation is necessary \cite{6},
not in order to tame quantum divergences but to fix quantum ambiguities,
approximation methods will be necessary in CQG as well. The challenge is to
show mathematically 
that truncations are really approximations, i.e. that there is 
some form of error control or convergence which to the best of our knowledge 
has not been established yet. 

Accordingly it is well motivated to have a fresh look at both programmes
and try to bring them into closer contact (see  \cite{Immirzi, relobs} for previous attempts). In \cite{3} we have tried 
to to give a compact description of both programmes useful for researchers 
from both communities in order to overcome differences of language. 
For ASQG practitioners we have laid out the basics of reduced phase space 
quantisation, relational observables, Hilbert space representations 
of the associated Weyl algebras supporting given 
Hamiltonian operators and the passage from the operator   
to the path integral formulation in
particular in the presence of gauge symmetries. For CQG practitioners 
we have reviewed the background field method in the presence of gauge 
symmetries, the effective (average) action, the Wetterich equation \cite{Wetterich:1992yh, Martin},
heat kernel techniques \cite{HK, HK1, HK2, HK3} for both signatures and truncation methods.

In application to GR one finds the following general features independent 
of the matter content of the system:\\ 
A. First, if the goal is to write 
the path integral in terms of the Einstein Hilbert action plus matter and 
further terms, then Lorentzian signature is selected. If one is 
content to write the theory just as some sort of path integral, then
the Euclidian version is also possible for the generating functional 
of Schwinger functions which are generated by the given Hamiltonian via 
Osterwalder-Schrader reconstruction \cite{7}.\\ 
B. Next, the path integral 
``measure'' deviates from the ``Lebesgue measure'' of the 
spacetime metric by several functions which 
depend on the chosen Hilbert space representation of the Weyl algebra, 
the determinant of the deWitt metric and as a result of the splitting 
of the metric into ADM variables. Some of these factors, but not all, 
can be absorbed 
into a ghost action which is related to the usual 
ghost action that results from the Fadeev Popov determinant 
as a result of gauge fixing the metric to some, say de Donder, 
gauge.\\
C. The Einstein Hilbert plus observable matter action is 
corrected by a ``reduction'' action 
which plays a role similar to a gauge fixing action but here is rather 
a fingerprint of how the gauge degrees of freedom have been absorbed 
into the observables of the theory in a way similar to the Higgs mechanism.
Accordingly, the true degrees of freedom contain $d(d+1)/2-d$ rather than 
$d(d+1)/2-2d$ observable gravity polarisations in $d$ spacetime 
dimensions as they have ``eaten'' $n$ scalar fields and have 
become ``Goldstone'' bosons. \\
D. Furthermore, the generating functional of Feynman distributions contains 
a current only for the observable $d(d+1)/2-d$ gravity polarisations,
not for all of them. For reasons of better comparability with the existing 
ASQG literature one can by hand augment the current by additional terms
to include also the unphysical $d$ gravity polarisations (which are 
integrated out in the correct non-augmented version by performing the lapse and 
shift integral) but one has to be 
careful in  the interpretation of the resulting effective action:
Its restriction to the observable part of the current is {\it not} the 
observable effective action that one is interested in, rather the two 
are related by a twisted combination of Legendre transforms and restriction
maps.\\
E. Finally, one has to add the cut-off action to define the effective 
average action. This, like the Einstein-Hilbert action comes with 
a factor of $i$ in the exponent in its Lorenzian version
and requires a different type of cut-off
functions than in the Euclidian signature case.\\ 
\\
In \cite{3} we have exemplified these general techniques for 
GR coupled to a very simple 
matter content as a showcase model, namely the Einstein Hilbert action
in $d$ spacetime dimensions minimally coupled to $d$ neutral, massless
Klein Gordon fields studied in 
\cite{GV} from a CQG perspective, in order to have a concrete example in 
mind. However, we only prepared this showcase model to make it ready for a 
ASQG treatment, the analysis itself was not carried out. 
This is the subject of the present paper. Thus we 
will write the concrete Wetterich equation for this model and derive 
the resulting flow equations in the lowest order truncation 
for the three coupling parameters on which this model depends. We show that 
there exist cut-off functions for which the required ``heat'' kernel
time integrals are finite. The Lorentzian heat kernel traces can 
otherwise be performed almost signature independently. In particular 
the so-called ``non-minimal'' term techniques \cite{8} can be copied 
literally. In this exploratory paper we ignore the effect of the 
ghost integral as a first step. This shows that the ASQG and CQG 
formalisms can be brought into contact, not only in principle but also 
very concretely. \\
\\
The architecture of this contribution is as follows:\\
\\

In section \ref{s2} we review the bare bones from \cite{3} necessary 
to be able to carry out the calculations of the present paper.

Section \ref{s3} contains the main results of this 
paper.  We formulate the non-perturbative Wetterich equation 
for this model and then truncate it to first few terms.
We compute the required heat kernel traces in that approximation 
paying attention to non-minimal terms. Next we perform the heat kernel 
time integrals with respect
to a new kind of cut-off that has to have different analytic properties than 
in the Euclidian signature case as what is relevant here are not Laplace
but Fourier transforms. As shown in \cite{3} (appendix C) these cut-off 
functions can also be used in the Euclidian regime. 
Finally we compute the flow equations for the 
three parameter
space of dimensionless 
couplings of the present model and study their solutions and fixed 
points. 
We are particularly interested in the $k\to 0$ limit of the 
dimensionful couplings which defines the effective action of actual interest.
A new feature of the Lorentzian flow is that at $k\not=0$ the 
couplings become generically complex valued. However, only the $k=0$ 
values of the couplings have physical meaning and thus the physically 
relevant or admissible trajectories are restricted 
to those for which all dimensionful
couplings are real valued at $k=0$.

In section \ref{s4} we summarise and conclude.

\section{The model and Lorentzian tools}
\label{s2}

The purpose of this section is to extract from \cite{3} the 
ingredients necessary in order to jump right away into the the ASQG 
treatment of the next section. 

\subsection{CQG derivation}
\label{s2.1}

The starting point is the classical action of the model
\be \label{2.1}
S=G_N^{-1}\; \int_M\; d^dx\; \sqrt{-\det(g)}\;[R[g]-2\Lambda-\frac{G_N}{2} 
S_{IJ}\; g^{\mu\nu}\;\phi^I_{,\mu}\;\phi^J_{,\nu}]
\ee
where $M$ is a d-dimensional manifold, necessarily diffeomorphic to 
$\mathbb{R}\times \sigma$ where $\sigma$ is a $(d-1)$ dimensional manifold
when $g$ is metric of Lorentzian signature 
and $(M,g)$ is a globally hyperbolic spacetime which is a necessary 
assumption in CQG \cite{9}. To avoid discussions about boundary terms we 
assume that $\sigma$ is compact without boundary, say a $d-$torus.
Furthermore, $R$ is the Ricci scalar, 
$G_N$ is Newton's constant and
$\Lambda$ is the cosmological constant. The matter content 
consists of $d$ neutral scalar fields $\phi^I,\;I=0,..,d-1$ which minimally 
couple to to the metric via a constant, real valued and positive
definite matrix $S$. 

The Legendre transform of the Lagrangian displayed in (\ref{2.1}) is singular
and leads to $2d$ first class constraints. There is a set of $d$ primary 
constraints $C^1_\mu$ which state that in the ADM formulation of (\ref{2.1}) 
the momenta conjugate to
lapse and shift functions are identically zero and these 
induce $d$ secondary constraints $C^2_\mu$ 
known as $d-1$ spatial diffeomorphism 
constraints and one Hamiltonian constraint \cite{10}.  
For the above model the unreduced phase space consists of the canonical 
pairs $(P_\mu, N^\mu), (p^{ab}, q_{ab}), (\pi_I,\phi^I)$ 
with $\mu=0,..,d-1,\; a=1,..,d-1$
where $N^0,N^a$ are lapse and shift functions respectively and $q_{ab}$ is the 
metric on $\sigma$.  The other variables denote the conjugate momenta.
In the reduced phase space approach one imposes $2d$ gauge fixing conditions,
on the configuration variables and solves the constraints for the 
$2d$ conjugate momenta. 
A convenient set of gauge conditions is that 
\be \label{2.2}
G_2^I:=\phi^I-k^I=0,\; G_1^\mu:=N^\mu-c^\mu=0 
\ee
where $k^I,c^\mu$ are functions on $M$ independent of the phase space 
variables subject to the condition that 
$\det(\partial k/\partial x)\not=0$. Accordingly we solve the constraints 
$C^2_\mu=0, C^1_\mu=P_\mu=0$
for $\pi_I, P_\mu$ which is trivial for $P_\mu$.
The secondary constraints can be solved algebraically 
for $\pi_I:=\pi^\ast_I$ \cite{3}. In this way the gauge degrees of freedom are 
identified as $(N^\mu,\phi^I),(P_\mu,\pi_I)$ while the true degrees of freedom 
are $(q_{ab}, p^{ab})$. 

To obtain the reduced Hamiltonian we note that the so called primary 
Hamiltonian for generally covariant systems is a  combination 
of constraints, in this case 
\be \label{2.2}
H_{{\sf primary}}=\int\;d^3x\; [v^\mu\; C^1_\mu+N^\mu\; C^2_\mu]
\ee 
where $C_0, C_a$ are the Hamiltonian and spatial diffeomorphism constraints.
Here $v^\mu$ are the velocities with respect to which the Legendre transform 
is singular. The gauge stability condition
imposes that the gauge fixing condition be 
preserved in time 
\be \label{2.3} 
\dot{G}_2^I
=\partial_t G_2^I+\{H_{{\sf primary}},G_2^I\}=    
\{H_{{\sf primary}},G_2^I\}-\dot{k}^I=0,\;
\dot{G}_1^\mu
=\partial_t G_1^I+\{H_{{\sf primary}},G_2^I\}=    
v^\mu-\dot{c}^\mu=0
\ee
which can be solved for $N^\mu:=N^\mu_\ast$ and 
$v^\mu:=v^\mu_\ast=\dot{c}^\mu=\dot{N}^\mu_\ast$. The reduced 
Hamiltonian is defined to be that function of the true degrees of
freedom only which generates the same equations of motion as the 
primary Hamiltonian when we restrict to the reduced phase space 
i.e. 
\be \label{2.4}
\{H,F\}=\{H_{{\sf primary}},F\}_\ast
\ee
where the subscript $\ast$ instructs to evaluate the Poisson bracket 
taken on the unreduced phase space and then 
freeze it to $P_\mu^\ast,\pi_I^\ast, N^\mu_\ast,
\phi^I_\ast, v^\mu_\ast$. Here $F$ is a function of $(q,p)$ only. 
The explicit expression for H is given in \cite{3}.
It is conservative, i.e. not explicitly time dependent iff  
$\kappa^I_\mu:=\partial k^I/\partial x^\mu$ is time $x^0$ independent. 
This is the simplest 
choice of gauge also adopted in \cite{3} and while not necessary 
simplies the formulae.

Equipped with the reduced phase space coordinatised by $(q,p)$, the 
reduced Hamiltonian and a preferred time direction $x^0$,
we can now quantise the system by imposing canonical 
commutation and $\ast$ relations in the usual way which gives rise to
a Weyl algebra $\mathfrak{A}$ 
for which we can pick Hilbert space representations. The 
cyclic representations correspond to states $\omega$ \cite{11}
with respect to which we can 
compute time ordered correlation functions (Feynman distributions) which
can be analytically continued in time 
due to the conservative nature of the Hamiltonian
(Schwinger distributions). These distributions are obtained from a 
generating functional $Z_s[f]$ which depends on currents $f^{ab}$ for $q_{ab}$ 
where $s=0,1$ corresponds to Euclidian and Lorentzian time respectively.

These generating functionals can be cast into a formal path integral
over the phase space spactime fields $(q,p)$  
by the usual methods and assumptions familiar from ordinary QFT
and depend on the Euclidian or Lorentzian phase space action induced 
by H respectively. Specifically
\be \label{2.5}
Z_s[f]=\int\; [dq\;dp]\; J_\omega[q]\;
e^{-\int\; dx^0\;(i\;<p,\dot{q}>+(-i)^s\;H[p,q])}
\;e^{i^s\;\int\;dx^0\;<f,q>}
\ee
where $<.,.>$ is the inner product on $\otimes^{d(d-1)/2}\; 
L_2(\sigma, d^{d-1}x)$ between symmetric twice contravariant tensor densities 
of weight one and twice symmetric covariant tensors and $J_\omega[q]$ is a 
functional that depends on the chosen state $\omega$ on $\mathfrak{A}$
\cite{3}. 

One now would like to perform the momentum integrals. This is difficult 
because $H$  involves a square root that originates from solving 
the constraints $C_\mu$ for the momenta $\pi_I$ on which they depend 
quadratically. Thus (\ref{2.5}) is not at all a simple Gaussian 
integral in $p$. In \cite{3} two possibilities for getting rid off 
the square root were  presented. The first method      
introduces a single auxiliary field $\lambda$ and is inspired by 
the observation that the critical point value of the 
function $\lambda\mapsto [\lambda\; h+\lambda^{-1}]/2$ is given by 
$\sqrt{h}$ so that in a saddle point approximation of the $\lambda$ integral
one obtains the square root. In \cite{3} the exact version of this saddle 
point argument is presented. It has the advantage that it works 
in principle for both signatures but it has the disadvantage 
of being spatially non-local involving 
the solution of PDE's that arise when integrating out $p$. We hope to 
come back to this method for the present model
in a future publication. Alternatively, there exists also 
scalar matter which avoids the square root from the outset \cite{12}
which would also be interesting to study. 

The second 
method follows the well known procedure \cite{13}
for unfolding a reduced phase space path integral to the unreduced phase 
space by introducing $\delta$ distributions for $G_2^\mu,C^2_\mu$     
and the determinant of the Dirac matrix $\Delta:=\{C^2,G_2\}$ which is very 
similar to the Lagrangian Fadeev-Popov method. Thus the 
path integral is extended to an integration also over $\phi^I,\pi_I,N^\mu$
where the integral over $N^\mu$ yields $\delta[C^2]$ and we keep 
$\delta[G^2]\; \det(\Delta)$ untouched. This enables to replace 
$H[q,p]$ by $-<\pi,\dot{\phi}>$ in (\ref{2.5}) because $H=-
<\pi_\ast,\dot{k}>$ at the price to 
augment the exponent by $-i\; <N,C^2>$ which installs the $\delta[C^2]$ 
distribution. We now see that for $s=0$ we run into trouble: The 
terms $<p,\dot{q}>,<\pi,\cdot{\phi}>$ come with a relative factor of 
$i$. Then carrying out the now Gaussian integrals over $\pi,p$ 
turns the exponent into the Einstein Hilbert action for {\it complex}
GR plus corrections. 

This forces us to work with $s=1$ from now on. The Gaussian integrals 
over $p,\pi$ can be 
performed which introduces a measure factor depending on $q,N$. 
We also carry out the integral over $\phi$ via 
$\delta[G_2]$ which replaces $\phi^I_{,\mu}$ by $k^I_{,\mu}=\kappa^I_\mu$. 
Then the 
integral only involves $q_{ab},N^\mu$. By switching integration variables
to $g_{\mu\nu}$ using the ADM relations $g_{00}=-N^2+q_{ab} N^a N^b,\;
g_{0a}=q_{ab} N^b,\; g_{ab} =q_{ab}$ we can write $[dq\;dN]=[dg] \; I[q,N]$
with the corresponding Jacobean $I[q,N]$ so that we end up with a functional
integral over Lorentzian signature spacetime metrics $g$, specifically 
\be \label{2.6}
\begin{aligned}
Z_1[f]&=\int\; [dg]\; J_\omega[g]\; 
e^{-i S_1[g]\;}
\;e^{i\int_M\; d^dx \; f^{ab} q_{ab}},\\
S_1[g]&=
\frac{1}{G_N}\int_M\;d^dx\sqrt{-\det(g)}[R[g]-2\Lambda-\frac{G_N}{2}
g^{\mu\nu}\;S_{IJ}\;\;k^I_{,\mu}\;k^J_{,\nu}]
\end{aligned}
\ee
where we have collected all measure factors that deviate from the exponential
of the displayed action into the function $J_\omega[g]$ \cite{3} 
written in terms of $g$ rather than $q,N$ which 
again depends on the state $\omega$. It can be written in terms
of a functional
determinant $J_\omega=\det(K_\omega)$, hence the second factor 
may be replaced by a ghost
integral if wanted \cite{3}
\be \label{2.7}
J_\omega[g]=
|\int [d\rho\;d\eta]\; e^{-i \int\; d^dx\; \eta^\mu \; 
[K_\omega]_\mu^I(g)\; \rho_I}|
\ee
where $K_\omega$ is the ghost matrix \cite{3} (we consider only the case 
$\kappa^I_\mu=$const. and $\kappa^I_0=\kappa_0 \delta^I_0$) 
\ba \label{2.7a}
[K_\omega]_\mu^I(g) 
&=& f_\omega(g)\{\frac{\delta_\mu^0}{N}[\kappa^I_0-N^a \kappa^I_a]
+\delta_\mu^a \kappa_a^I\},\; 
\nonumber\\
f_\omega(g) &=&
|-\det(g)|^{2/d-(d+1)/8}\; |\det(q)|^{2/d+(d+1)/4}\;
\exp(h_\omega/d)
\nonumber\\
h_\omega(x^0,\vec{x})&=& 
[\delta(x^0,\infty)+\delta(x^0,-\infty)]\;
I_\omega(x^0,\vec{x}),\;
\Omega_\omega[q(x^0)]\nonumber \\&=&:\exp(\int\; d^{d-1}x\; I_\omega(x^0,\vec{x}))
\ea
and $q,N$ are to be expressed in terms of $g$ using the above ADM relations.
Here $\Omega_\omega[q(x^0)]$ is the cyclic vector of the GNS data
at fixed $x^0$ underlying the state $\omega$, written 
in the configuration presentation. It
contributes only for $x^0=\pm\infty$. More details about $I_\omega$ can 
be found in \cite{3}. The $k^I$ dependent ``reduction term'' replaces the
usual gauge fixing term but is logically independent of it. 
   
We will now drop the index ``1'' in $Z_1$ which reminds of the fact that we 
are dealing with Lorentzian signature. The fact that $Z[f]$ depends on 
$f^{ab}$ only and not on the full $f^{\mu\nu}$ reminds us of the fact 
that we are dealing with a reduced phase space formulation and thus 
only correlation functions of the observable spacetime field $q_{ab}(x)$ are 
accessible. In principle one can integrate out lapse and shift in 
(\ref{2.6}) to obtain a path integral just for $q$. It is remarkable 
that the only effect on the exponential of the action due to 
the reduced phase space formulation is that the 
Einstein-Hilbert action is corrected by the non-covariant ``reduction'' 
term involving 
$k^I$. We choose $k^I_{,\mu}=:\kappa^I_\mu$ to be constant and thus 
can abbreviate the constant matrix 
$\kappa_{\mu\nu}:=S_{IJ} \kappa^I_\mu \kappa^J_\nu$ which introduces 
$d(d+1)/2$ coupling constants. The other non-covariant term is the measure 
factor $J_\omega[g]$. Both non-covariances again remind us of the fact that
we have fixed a certain gauge and all statements about correlators have 
to be translated into each other by the corresponding spacetime 
diffeomorphisms when switching gauges. Note that the gauge chosen ties 
the spacetime coordinates to a dynamical reference field $\phi^I$, therefore 
the coordinates become observable and in that sense the description is 
gauge independent but dependent on the interpretation of the coordinates.
See \cite{3} for more details.  

\subsection{ASQG treatment}
\label{s2.2}

To set up the system (\ref{2.6}) for an ASQG treatment we formally extend 
the current to include also $u^\mu:=f^{0\mu}$ components which we remind 
us of by switching notation from $f$ to $F$ where $F^{0\mu}=u^\mu,\;
F^{ab}=f^{ab}$ and $Z\to Z'$. 
Then we define the effective action in the usual way 
by 
\be \label{2.8}
C'[F]=i^{-1}\; \ln(Z'[F]),\; \Gamma'[\hat{g}]:=
[L\cdot C'][\hat{g}]:={\sf extr}_F (<F,\hat{g}>-C'[F])
\ee
where $L$ denotes the Legendre transform.
Note however that while $Z[f]=[R\circ Z'][f]:=
Z'[F]_{u=0}, \; C[f]=C[F]_{u=0}$ are just related by restriction $R$ 
it is 
{\it not} true that $\Gamma[\hat{q}]=\Gamma'[\hat{g}]_{\hat{N}^\mu=0}$. 
Rather \cite{3} 
\be \label{2.9}
\Gamma=L\circ R\circ L^{-1} \cdot \Gamma' 
\ee
which we need to keep in mind because what we are interested in is 
$\Gamma$ and not $\Gamma'$. In QFT one considers a well defined $\Gamma$
a complete solution of the theory.

In ASQG one works with the background field method. Thus in 
(\ref{2.6}) we replace $g$ everywhere by $\bar{g}+h$ and $[dg]$ by 
$[dh]$ except in $<F,g>$ which is replaced by $<F,h>$. The resulting 
generating functional is denoted by $\bar{Z}'[F,\bar{g}]$ and corresponding 
$\bar{C}'[F,\bar{g}],\; \bar{\Gamma}'[\hat{g},\bar{g}]$. As is well 
known, we recover the background independent effective 
action by 
\be \label{2.10}
\Gamma'[\hat{g}]:=\bar{\Gamma}'[\hat{g}';\bar{g}]_{\hat{g}'=0,\;\bar{g}:=\hat{g}}
\ee
Finally we introduce the effective average action \cite{Wetterich:1992yh, Wett2, Wett3} through the chain 
of relations  
\ba \label{2.11}
\bar{Z}'_k[F,\bar{g}] &=& \int\; [dh] \; J_\omega[\bar{g}+h]\; 
e^{-i S[\bar{g}+h]}\; e^{i<F,h>} e^{-i\frac{1}{2} <h, R_k(\bar{g})\cdot h>},\;\;
\nonumber\\
\bar{C}'_k[F,\bar{g}]&=& i^{-1}\; \ln(\bar{Z}'[F,\bar{g}]),\;
\nonumber\\
\bar{\Gamma}'_k[\hat{g},\bar{g}] &=& {\sf extr}_F\;(<F,\hat{g}>-
C'_k(F,\bar{g}))-\frac{1}{2}<\hat{g},R_k(\bar{g})\cdot \hat{g}>
\ea
where $k\to R_k(\bar{g})$ is a 1-parameter family 
of background dependent integral kernels which only depends
on the background d'Alembertian. In the Euclidian signature 
case it intuitively corresponds to a suppressing kernel for Euclidian
momenta below $k$, in the Lorentzian case suppressing is replaced by 
oscillations although it is clear that null modes cannot be tamed 
like this. Therefore we will only take over one of the properties of 
$R_k$ from the Euclidian case namely $R_k=0$ for $k=0$ while 
we adapt the other properties of $R_k$ to Lorentzian signature 
further below. This ensures 
that $\bar{\Gamma}'(\hat{g},\bar{g})=\bar{\Gamma}'_0(\hat{g},\bar{g})$ so that 
the object of actual interest (\ref{2.9}) is available from 
$\bar{\Gamma}'_k(\hat{g},\bar{g})$ through the chain of relations displayed.

The importance of $\bar{\Gamma}'_k(\hat{g},\bar{g})$ lies in the fact that 
it obeys the Lorentzian version of the {\it Wetterich equation}  
\be \label{2.12}
k\;\partial_k \;\bar{\Gamma}'_k[\hat{g},\bar{g}]
=\frac{1}{2i}\; {\sf Tr}([R_k(\bar{g})+
\bar{\Gamma}^{\prime(2)}_k(\hat{g},\bar{g})]^{-1}\; [k\;\partial_k R_k(\bar{g})]),\;
\bar{\Gamma}^{\prime(2)}[\hat{g},\bar{g}]:=
\frac{\delta^2 \bar{\Gamma}^{\prime(2)}[\hat{g},\bar{g}]}{
\delta\hat{g}\otimes \delta\hat{g}}
\ee
This integro functional differential equation is an exact and 
non-perturbative identity and can be used to construct a well defined 
$\Gamma$ rather than using its ill defined expression (\ref{2.6}).

To solve (\ref{2.12}) exactly one Taylor expands both the l.h.s. and r.h.s 
in powers of $\hat{g}$ and compares coefficients. 
This gives an infinite iterative hierarchy of relations because 
(\ref{2.12}) connects the Taylor coefficients of order $n$ to those of
order $n+2$. In practice one has to truncate at some finite order 
$T$ of Taylor coefficients on the l.h.s. that we want to take into account.
Often one just considers $T=0$. 

To actually compute the traces on the r.h.s. of the 
Wetterich equation for the $k\partial_k$ derivative 
of the $\hat{g}$ independent 
N-th order Taylor coefficients 
$T^N_k(\bar{g})=\bar{\Gamma}^{\prime(N)}_k(\hat{g},\bar{g})_{\hat{g}=0}$
one notices that these can be written, (we do not display 
the dependendence on $\bar{g}$) 
\be \label{2.13}
{\sf Tr}([P_k+R_k+U_k]^{-1}\; [k \partial_k R_k]\;
[P_k+R_k+U_k]^{-1}\; V^1_k\;[P_k+R_k+U_k]^{-1}\;..\;
[P+R_k+U_k]^{-1}\; V^M_k)
\ee
$M=0,..,T$ where $T^2_k=P+U_k$ has been split into a term $P_k$ which depends 
on $\bar{g}$ only through the background d'Alembertian 
$\overline{\Box}=\bar{g}^{\mu\nu}\;\bar{\nabla}_\mu\;\bar{\nabla}_\nu$.
The operators $U_k, V^I_k$ are not necessarily 
such ``minimal''operators and can have general dependence on 
$\bar{g}$. Then we expand 
$(1+P_k^{-1} [R_k+U_k])^{-1}$ 
into a geometric 
series. In practice 
one has to truncate that series at some order $S$. This basically 
replaces (\ref{2.13}) to the effect that 
$R_k+U_k$
in the denominator is dropped 
and the $V^I_k$ are replaced by other in general non-minimal operators.
Then every minimal operator factor in (\ref{2.13}) is replaced 
via the spectral theorem by 
\be 
\label{2.14}
O_k(\overline{\Box})=\int_{-\infty}^\infty\; dt\; \hat{O}_k(t)\; 
H_t,\;\; H_t:=e^{it\overline{\Box}}
\ee
where $\hat{O}_k(t)$ is the Fourier transform of $O_k(z)$ and $H_t$ the 
``heat'' (better: Schr\"odinger) kernel \cite{HK, HK1, HK2, HK3}. It follows that we are interested 
in the traces 
\be \label{2.15} 
\begin{aligned}
&\int \; d^{M+1} t\; \prod_{I=0}^M\; \hat{O}_k^I(t_I)\;
{\sf Tr}(V^1_k(t_0)
V^2_k(t_0+t_1)\;..\;
V^M_k(t_0+..+t_{M-1})\; H_{t_0+..+t_M}),\;\\
&V^I_k(s)=H_s \; V^I_k\; H_{-s}
\end{aligned}
\ee
The $V^I_k(s)$ can be Taylor expanded with respect to $s$ which 
we truncate at some order $R$. This replaces the $V^I_k(s)$ in 
(\ref{2.15}) by other 
non-minimal but $s-$independent operators $V^I_k$ and introduces a polynomial 
in the $t_0,..,t_M$ so that we are interested in 
\be \label{2.16} 
\int \; d^M t\; \prod_{I=0}^M\; \hat{O}_k^I(t_I)\;
{\sf Pol}(t_1,..,t_M)\; {\sf Tr}(V^1_k\;..\;V^M_k\; H_{t_0+..+t_M})
\ee
The remaining trace can be computed using heat kernel techniques as 
detailed below.\footnote{In \cite{Bonanno:2000yp, Mazza:2001bp, Bonanno:2004sy, Bonanno:2004sy.1} different variants of the proper time flow equation were presented and analysed.}  

Finally one parametrises 
$\bar{\Gamma}'_k(\hat{g},\bar{g})$
in terms of a suitable basis of $\bar{\Gamma}'_\alpha(\hat{g},\bar{g})$ of 
``actions'' that come with dimensionful ``couplings'' $C_{k,\alpha}$ where 
$\alpha$ runs through a countable index set. In practice we truncate 
the number $A$ of $\alpha$ that we retain, subordinate to the truncation 
parameters $R,S,T$ above in such a way that we obtain a
closed autonomous system of first order 
ODE's for the $C_{k,\alpha}$. One factors off the dimension 
of those couplings and obtains an autonomous closed system of first order 
ODE's for dimension free couplings 
$c_{k,\alpha}=k^{-d_\alpha} C_{k,\alpha}$, specifically 
$k\partial_k c_k=:\beta(c_k)$. Consider a UV ($k\to\infty$)
fixed point $c^\ast$ of this 
flow i.e. $\beta(c^\ast)=0$. It is called a predictive fixed point when 
all but a finite (and $T,S,R,A$ independent) number of the $c_\alpha$ must 
be fine tuned to the fixed point values $c^\ast_\alpha$ 
in order that the fixed point is reached. The remaining parameters are the 
relevant parameters that need to be measured while the fine tuned ones 
are predictions of that fixed point. 

From the point of view of CQG this is the only purpose of going all the 
way through these steps because it offers a way to define the theory 
(\ref{2.9}) that we are interested in, provided that the limit 
$k\to 0$ of $\Gamma'_k(\hat{g},\bar{g})$ can be taken. Thus 
the $k\to \infty$ limit of $c_k$ and the      
the $k\to 0$ limit of $C_k$ must co-exist for this particular fixed point. 

\subsection{Lorentzian heat kernel expansion, time integrals and 
cut-off functions}
\label{s2.3}       

The Lorentzian heat kernel on the Lorentzian 
spacetime $(M,\bar{g})$ is the solution to the initial value problem
\be \label{2.17}
[\partial_t-i\overline{\Box}]\;H_t(x,y)=0,\;
H_0(x,y)=\delta(x,y)
\ee
The heat kernel time $t$ has nothing to do with the time coordinate 
$x^0$. For Minkowski space $(M,\bar{g})=(\mathbb{R}^d,\eta)$ one finds the 
unique solution 
\be \label{2.18}
H_t(x,y)=[4\pi |t|]^{-d/2}\; e^{i\frac{\pi}{4}{\sf sgn}(t)[2-d]}\;
e^{\frac{i}{2t}\;\sigma(x,y)} \;
\sigma(x,y)=\frac{1}{2}\eta_{\mu\nu}(x-y)^\mu\;(x-y)^\nu           
\ee
On general $(M,\bar{g})$ one generalises (\ref{2.18}) to
\be \label{2.19}
H_t(x,y)=[4\pi |t|]^{-d/2}\; e^{i\frac{\pi}{4}{\sf sgn}(t)[2-d]}\;
e^{\frac{i}{2t}\;\sigma(x,y)} \;\Omega_t(x,y)
\ee
where $\sigma(x,y)$ is called the {\it Synge world function} i.e. the 
{\it signed} square of the geodesic distance between $x,y$ (positive, 
negative, zero when the geodesic between $x,y$ is spacelike, timelike or 
null) which are assumed to lie in a convex normal neighbourhood. It satisfies 
the master equation
\be \label{2.20}
\bar{g}^{\mu\nu}(x)\;
[\bar{\nabla}^x_\mu \sigma](x,y)\;
[\bar{\nabla}^x_\nu \sigma](x,y)\;=2\;\sigma(x,y),\; \sigma(x,x)
\ee
This equation, which is remarkably signature insensitive, allows to compute 
the coincidence limit $y\to x$ of all covariant derivatives of $\sigma$  
in terms of the curvature tensor of $\bar{g}$. 

One now plugs the Ansatz (\ref{2.19}) into (\ref{2.17}) and obtains 
a PDE for $\Omega_t$ subject to the initial condition $\Omega_0(x,x)=1$ 
as the prefactor in (\ref{2.19}) already produces $\delta(x,y)$ at $t=0$.
To turn that PDE into a system of algebraic equations one first expands 
$\Omega_t$ with respect to the heat kernel time $t$
\be \label{2.21}
\Omega_t(x,y)=\sum_{k=0}^\infty\;(it)^k\; \Omega_k(x,y)
\ee
The factors of $i$ are chosen such that the $\Omega_k$ are real valued also 
in Lorentzian signature. One finds (metric coefficients and derivatives at
$x$)
\be \label{2.22}
[\frac{\overline{\Box}-d}{2t}+k]\Omega_k+\bar{g}^{\mu\nu}
[\bar{\nabla}_\mu \Omega_k]\;[\bar{\nabla}_\nu \sigma]
-[\overline{\Box}\Omega_{k-1}]=0
\ee
with $\Omega_{-1}(x,y)\equiv 0,\; \Omega_{k=0}(x,x)=1$. Then we perform a 
coincidence limit expansion 
\be \label{2.23}
\Omega_k(x,y)=\sum_{l=0}^\infty\; \frac{1}{l!}\; 
[\Omega_{k,l}]^{\mu_1..\mu_l}(x)\; 
[\bar{\nabla}^x_{\mu_1} \sigma](x,y)..[\bar{\nabla}^x_{\mu_l} \sigma](x,y)
\ee
Plugging (\ref{2.23}) into (\ref{2.22}) allows to compute all completely 
symmetric tensors 
$\Omega_{k,l}(x)$ algebraically just using the master equation. 
By the same methods also the evaluation of non-minimal derivative 
operators on the heat kernel can be evaluated algebraically. All that is 
needed is the master equation. Since these relations do not depend on the 
signature we can transfer without change literally all the listed 
expressions for $\Omega_{k,l}$ from the Euclidian to the Lorentzian 
regime. See \cite{8,3} and references therein for more details.

Once all of this has been done one takes the trace in which consists
in evaluating 
\be \label{2.24}
\int\; d^dx\; [V^1_k(x)\;..\; V^M_k(x) H_s(x,y)]_{y\to x}
\ee
which is why the coincidence limit is important and why the assumption
of $y$ to lie in a convex normal neighbourhood of $x$ is justified. The 
notation in (\ref{2.24}) means that that the $V^I_k$ are considered as 
differential operators that act on the $x$ dependence of the heat 
kernel before taking the coincidence limit. These integrals return 
expressions depending on the background metric, the 
curvature tensor, derivatives thereof and thus allow for an 
unambiguous comparison of coefficients when computing the $\beta$ 
functions of the flow provided $\bar{g}$ is kept arbitrary.

The final step consists in computing the integrals over the heat kernel 
times and it is at this point where the choice of the cut-off function
$R_k$ becomes crucial. The following is a possible choice introduced 
in \cite{3} which serves as a proof of principle that the Lorentzian heat 
kernel time integrals converge for suitable $R_k$ but it is only motivated 
by the mathematical convergence property and not a physical principle.

From the discussion above it is clear that the operators $P_k^{-1},\;
R_k$ play a fundamental role. They need to be expressed in terms 
of the heat kernel. As a typical example we consider 
$P_k(\overline{\Box})=\Lambda_k+B\;\overline{\Box}$ there 
$B$ is independent of $k$. Then with $C_k=\Lambda_k/B$ 
\be \label{2.27}
P_k^{-1}=B^{-1}\;[\overline{\Box}+C_k]^{-1}=
-\frac{i}{B}\; [\int_0^\infty\;dt\; 
e^{it\overline{\Box}-t \epsilon}]_{\epsilon\to -i C_k}
\ee
is like a Schwinger proper time integral involving 
the heat kernel where it is understood that one 
performs the integral at $\epsilon>0$ and then analytically continues 
$\epsilon\to -i C_k$ at the end. Note that the heat kernel time 
integral in (\ref{2.27}) is confined to the positive real axis.

Furthermore we pick 
\be \label{2.28}
R_k(z)=f_k\; k^2 \; r(z/k^2),\;
r(y)=\int_0^\infty\;dt\; e^{-t^2-t^{-2}}\; e^{it y}
\ee
where $f_k$ is a function of the couplings which is equips $R_k$ with 
the correct physical dimension and which in practice increases the 
non-linearity of the flow.
Thus the Fourier transform $\hat{r}$ has rapid decrease at $t=0,+\infty$
and smoothly joins the constant function $\hat{r}(t)\equiv 0,\; t\le 0$. 
Thus no boundary terms arise when integrating by parts.

The reason for this choice is the following:
the heat kernel time integrals (\ref{2.16}) involve the heat kernel
$H_s,\; s=t_0+..+t_M$ and the heat kernel itself contains $|s|^{-d/2}$
as a prefactor. If the heat kernel time integrals had support also 
on the negative real axis then there would be multiple configurations 
of $t_0,..,t_M$ which yield poles $s=0$ and 
none of the heat kernel time integrals would converge. This cannot 
happen when all heat kernel times are positive. Furthermore, all
integrals that appear contain at least the factor of 
$k\partial_k R_k$  corresponding 
to the $t_0$ integral in (\ref{2.16}) producing a $e^{-t_0^2-t_0^{-2}}$ 
factor. The basic estimate $s^{-d/2}\le t_0^{-d/2}$ shows that therefore the 
required integrals converges absolutely.

The price to pay is that these integrals become complex valued making
the flow of the couplings complex valued. This in principle doubles the 
number of real couplings. However, we are not interested in all complex 
trajectories but only the {\it admissible} ones. These are those with 
the property that the dimensionful couplings have a real valued 
$k\to 0$ limit when they exist. This is a form of fine tuning and 
halves the number of initial conditions (trajectories) of the flow 
so that effectively one is dealing with the same dimensionality       
of the flow as in the Euclidian case.
Note that in the Euclidian signature case the heat kernel times
are automatically confined to the real axis because one is dealing with
the one sided Laplace transform rather than the Fourier transform.

In the literature on Euclidian signature ASQG multiple heat kernel 
time integrals, at least when only minimal operators are 
involved, are avoided by assuming that a given function $F(y),\;y\ge 0$ is in 
the image of the Laplace transform, i.e. that there exists 
$\hat{F}(t)$ such that $F(y)=\int_0^\infty\; dt\; \hat{F}(t)\; e^{-yt}$ 
Then it follows that for $n\in \mathbb{Z}$  
\be \label{2.28a}
\int_0^\infty\; dt\; t^n\; \hat{F}(t)=
\theta(n)\;(-1)^n\; F^{(n)}(0)
+\theta(-n)\; \frac{1}{(|n|-1)!}\int_0^\infty\; dy\; y^{|n|-1}\; F(y)
\ee
so that one never needs to know $\hat{F}$. In \cite{3} we show that the 
existence of $\hat{F}$ for commonly used cut-off functions is by no 
means secured. This is the reason why we start here with given $\hat{F}$
whose existence is secured. Therefore relations of the type (\ref{2.28a})
are of little practical use as we only 
know $\hat{F}$ explicitly rather than $F$. 
In \cite{3} it is shown that (\ref{2.28}) is also a valid choice in the 
Euclidian regime (with $iy\to -y<0$). 

To solve the Wetterich equation one typically starts with the 
1-loop background effective action as an Ansatz which is given by
\be \label{2.28b}
\bar{\Gamma}'(\hat{g},\bar{g})=S(\hat{g}+\bar{g})+
\frac{1}{2i}\;[
{\sf Tr}(\ln[S^{(2)}))-2\;{\sf Tr}(\ln[K_\omega]))](\hat{g}+\bar{g})
\ee
where $K_\omega$ is the ghost matrix and then makes the couplings of the 
various terms in this expression dependent on $k$. Here $S$ already contains 
the reduction term which in some sense replaces 
the gauge fixing term in the usual 
treatment and the logarithmic term replaces the ghost action flow 
in the usual treatment.           
     
\section{ASQG analysis of the model}
\label{s3}

In this exploratory paper we will be content with lowest order truncations 
$R,S,T,A$ in order to gain experience.\footnote{See  for instance Refs. \cite{14, 14.1} for an Euclidian ASQG treatment of higher order truncations in gravity and gravity-coupled matter systems.} In particular we will ignore 
the effect of the ghost matrix $K_\omega$, 
consider only the zeroth order $T=0$ Taylor 
expansion of the Wetterich equation with respect to $\hat{g}$, 
expand the geometric series involved 
in the trace of the Wetterich equation only up to $S=2$ in the 
non-minimal terms, 
keep only the zeroth order $R=0$ with respect to $s$ in the 
``heat kernel evolved'' non-minimal operators $H_s V^I_k H_{-s}$ and 
finally truncate the flow of actions with respect to an $A=3$ dimensional
space of dimensionful couplings corresponding to Newton's constant $G_N$,
the cosmological constant $\Lambda$ and $\kappa$ where we specialise 
the gauge $\phi^I=k^I$ such that $\kappa^I_\mu=k^I_{,\mu}=$ const.
and such that $\kappa_{\mu\nu}:=S_{IJ} \kappa^I_\mu \kappa^J_\nu=
\kappa \delta_{\mu\nu}$ with $\kappa>0$. This means 
that we study the concrete problem 
\be \label{3.1}
k\partial_k \gamma_k=\frac{1}{2i} 
\text{Tr}
\{
[k\partial_k R_k]\;P_k^{-1}
(1-[(R_k+U_k)\; P_k^{-1}]+[((R_k+U_k)\; P_k^{-1}]^2)
\}
\ee
where 
\be \label{3.2}
\gamma_k:=\bar{\Gamma}_k^{\prime(0)}(0,\bar{g}),\;
\bar{\Gamma}_k^{\prime(2)}(0,\bar{g})=:P_k+U_k,\;
\ee
and $P_k=G_{N,k}^{-1}[\Lambda_k+B\overline{\Box}]$ collects all terms 
which depend only on $\overline{\Box}$ (minimal terms) while 
$R_k=G_{N,k}^{-1}\;k^2 r(z/k^2)$ with $r$ as in the previous section.

The Ansatz is then given by
\be \label{3.3}
\bar{\Gamma}'_k(\hat{g},\bar{g}):=
\{\frac{1}{G_{N,k}}\int\;d^dx\;[-\det(g)]^{1/2}[R[g]-2\Lambda_k
-\frac{\kappa_k G_{N,k}}{2} g^{\mu\nu}\; \delta_{\mu\nu}]\}_{g=\bar{g}+\hat{g}}
\ee
A similar Ansatz was used also in \cite{relobs}, 
where a preliminary study of relational observables in ASQG was carried out.

In what follows we will now go step by step through the ASQG treatment 
of the concrete truncation given above discarding all terms on the 
r.h.s. of the Wetterich equation that are 
not of the form (\ref{3.6}).

\subsection{Evaluation of the heat kernel traces}
\label{s3.1}

 In order to evaluate \eqref{3.2} via the heat kernel traces, the first step is to compute the  Hessian
\ba \label{3.4}
[\bar{\Gamma}^{\prime(2)}]^{\mu\nu}\;_{\rho\sigma}&=&
\frac{[-\det(\bar g)]^{1/2}}{G_{N,k}} \left((\bar\Box+2\Lambda_k) K^{\mu \nu}{}_{\rho \sigma} + U_k^{\mu \nu}{}_{\rho \sigma}\right)\;,\\
K^{\mu \nu}{}_{\rho \sigma}&=&\frac{1}{4}\bar\delta^{\mu}_ {\rho}  \bar\delta^{\nu}_ {\sigma}+\frac{1}{4}\bar\delta^{\nu}_ {\rho}\bar\delta^{\mu}_ {\sigma}-\frac{1}{2}\bar g^{\mu \nu} \bar g_{\rho \sigma}\;,\label{3.5}
\\
U_k^{\mu \nu}{}_{\rho \sigma}&=&\frac{1}{2}\left( \bar D^{(\mu}\bar D^{\nu)} \bar g_{\rho \sigma} + \bar D_{(\rho}\bar D_{\sigma)} \bar g^{\mu \nu} - \bar D^{(\mu}\bar D_\alpha \bar\delta^{\alpha)}_{\sigma}\bar\delta^{\nu}_{\rho}-\bar D^{(\mu}\bar D_\alpha \bar\delta^{\alpha)}_{\rho}\bar\delta^{\nu}_{\sigma}\right)\nonumber\\
\label{3.6}
&&+\frac{1}{2}(\bar R^{\mu}{}_\rho{}^\nu{}_\sigma+\bar R^{\mu}{}_\sigma{}^\nu{}_\rho)+\frac{1}{4}\left(\bar\delta_\rho ^\mu \bar R^\nu{}_\sigma + \bar\delta_\sigma ^\mu \bar R^\nu{}_\rho +\bar \delta_\rho ^\nu \bar R^\mu{}_\sigma+\bar \delta_\sigma ^\nu \bar R^\mu{}_\rho\right)\\
&&-\frac{1}{2}\left(\bar g^{\mu \nu }\bar R_{\rho \sigma}+ \bar g_{\rho\sigma} \bar R^{\mu \nu}\right)-\frac{1}{4}\bar R\left(\bar\delta^\mu_{\rho}\bar\delta^\nu_{\sigma}+\bar\delta^\nu_{\rho}\bar\delta^\mu_{\sigma}-\bar g^{\mu \nu} \bar g_{\rho \sigma}\right) \nonumber\\
&&+ \frac{1}{2} \Lambda_k\bar g^{\mu \nu}\bar g_{\rho \sigma}-\frac{ G_{N,k}\kappa_k}{4}\left(\delta^{\mu}_ {\rho} \bar g^{\nu}_{\sigma}+\delta^{\nu}_ {\rho} \bar g^{\mu}_{\sigma}-(\delta^{\mu \nu}\bar g_{\rho \sigma}+ \delta_{\rho \sigma } \bar g ^{\mu \nu }\right.)\nonumber\\&&\left.\qquad-\frac{1}{2}\delta^{\alpha }_\alpha\left(\bar\delta^{\mu}_ {\rho}  \bar\delta^{\nu}_ {\sigma}+\bar\delta^{\nu}_ {\rho}\bar\delta^{\mu}_ {\sigma}-\bar g^{\mu \nu} \bar g_{\rho \sigma}\right)\right)\nonumber\;,
\ea
where we distinguished between 
$\bar \delta_\nu^\rho = \bar g_{\mu \nu}\bar g^{\mu \rho}$ to be the background gravitational metric and  $ \delta_{\mu \nu}$ to be the matrix involved in the matter contribution having set $\kappa_{\mu \nu,k} = \kappa_k \delta_{\mu \nu}$. The indexes are raised or lowered through the background metric $\bar g$. Furthermore, we can now identify the structure of $P_k$ given in \eqref{2.27} with 
$
B = 1
$
and
$C_k = 2\Lambda_k$.

Making the Ansatz that the regulator $R_k$ has the tensorial structure 
\be\label{3.7}
R_k ^{\mu \nu}{}_{ \rho \sigma}= [-\det(\bar g)]^{1/2} G_{N,k}^{-1} K^{\mu \nu}{}_{ \rho \sigma} k^2 r(z/k^2)
\ee
simplifies considerably the computations.\footnote{Remember that $R_k^{\mu \nu \rho\sigma}$ introduced in \eqref{2.11} is the integral kernel of $\hat g_{\mu \nu}$ and $\hat g_{\rho \sigma}$. Here, for convenience, we lowered two indexes with the background metric $\bar g$.} Effectively, one is left with computing the suitable trace of products of $K^{-1} U_k$ up to the order established in the expansion (\ref{3.1}), where the matrix $K^{-1}$ is the inverse of $K$ defined in \eqref{3.5}:
\be
(K^{-1})^{\mu \nu}{}_ {\rho \sigma} = \bar \delta^\mu_\rho \bar \delta^\nu_\sigma + \bar \delta^\nu_\rho \bar \delta^\mu_\sigma- \frac{1}{d-1} \bar g^{\mu \nu } \bar g_{\rho \sigma}\;.
\ee

From this point onwards, 
all the evaluations have been specialized to $d=4$ spacetime dimensions. A first observation we make is that the tensorial trace of the product between $K^{-1}$ and the terms in the last line in \eqref{3.6}, containing the contributions coming from the matter, gives in general dimension a contribution proportional to
$
\kappa_k G_{N,k}\delta^\alpha_\alpha (8-2d-4d^3+d^4)
$
which is exactly vanishing in $d=4$. Also at order $[K^{-1}U_k]^2$ this 
results in  a coefficient proportional to 
$\kappa_k G_{N,k}\delta_{\mu \nu}g^{\mu \nu}(48 - 84 d + 18 d^2) $, 
vanishing in $d=4$.  The implication of this, is that the coupling 
constant $\kappa_k$ is not flowing and the flow of the gravitational 
couplings $\Lambda_k$ and $G_{N,k}$ completely decouples from the matter 
content. The additional matter term comes from 
the phase space reduction and indicates how scalar field degrees 
of freedom are transformed into metrical ones.
However, at this level of the truncation, it neither explicitly 
contributes to the flow of the couplings related to the physical 
degrees of freedom, nor is it itself affected by their running.

 Now we are able to explicitly write down term by term the heat kernel traces in (\ref{3.1}). As customary in RG analysis we switch to dimensionless variables, i.e. concretely,
 \be\label{3.8}
y=z/k^2, \qquad
\Lambda_k = \lambda_kk^2, \qquad 
G_{N,k} = \frac{g_k}{k^2}, \qquad
U_k = u_k k^2, \qquad
 \eta_N = \frac{k \partial_k g_k}{g_k}\;,
 \ee
 where $\eta_N$ stands for the anomalous dimension of the dimensionless Newton's coupling.
 
 At zero-th order one is left with
 \be\label{3.9}
 \begin{aligned}
& \text{Tr} \frac{(2-\eta_N)r(y)- 2 i y  r'(y)}{y+2\lambda_k} 
 =\\&\qquad=
 -i \text{Tr} \int_0^\infty dt_1 dt_2 e^{iy(t_1+t_2)}e^{-\epsilon t_1}\big((2-\eta_N)e^{-t_2^2-t_2^{-2}}+ 2 \; \frac{d}{dt_2}(e^{-t_2^2-t_2^{-2}})\big)\Bigg|_{\epsilon\to -2 i \lambda_k}
  \end{aligned}
 \ee
 At first order
 \ba\label{3.10}
&& \text{Tr}\frac{\big(2-\eta_N)r(y)- 2i y  r'(y)\big)(r(y)+u_k)}{(y+2\lambda_k)^2}
=-\int_0^\infty dt_1 dt_2 dt_3   e^{iy(t_1+t_2+t_3)}e^{-\epsilon (t_1+t_3)}\cdot
\\
&&\qquad\qquad\big((2-\eta_N)e^{-t_2^2-t_2^{-2}}+ 2 \; \frac{d}{dt_2}(e^{-t_2^2-t_2^{-2}})\big)(\int_0^\infty dt_4 e^{iyt_4}e^{-t_4^2-t_4^{-2}} + u_k)\Bigg|_{\epsilon\to -2 i \lambda_k}\nonumber
\ea

At second order the term has an analogous structure with up to 6 integrations in proper time variables $t_i$, which we do not report for sake of readability. 

In order to perform the traces  we specialize to $d=4$, note that
the integrals only involve positive values of $t$ and 
we exploit the heat kernel formula in general manifolds \eqref{2.19}. 
In particular the $\Omega_k$ up to the order we are 
interested in can be found in the literature \cite{HK, HK1, HK2, HK3, 8, 8.1, 8.2}
\ba\label{3.11}
	H &=&\frac{(-i)}{( 4\pi t)^{d/2} }(\Omega_0 +i t\,\Omega_1)\,,\\
	\label{3.12}
	H_{(\mu \nu)} (x,y) &=& \frac{(-i)}{( 4\pi t)^{d/2} }\left(-\frac{1}{2t}g_{\mu \nu}\Omega_0
	-\frac{i}{2}g_{\mu \nu}\Omega_1+i\bar D_{(\mu}\bar D_{\nu)}\Omega_0 \right)\,.
\ea
where $H$ is to be applied for minimal operators, while  
$H_{(\mu \nu)}$ for non-minimal operators involving two derivatives. The $ \Omega $'s are given by
\ba\label{3.13}
	&\Omega_0 =1 \,, \hspace{2cm}
	\Omega_1 = \frac{1}{6}\bar R  \,,\hspace{2cm}
	\bar{D}_{(\mu}\bar{D}_{\nu)} \Omega_0=\frac{1}{6}\bar R_{\mu \nu} \;.
\ea

At zero-th order one is left only  with the minimal term and the trace reduces to a proper time integral of the form
\be
\begin{aligned}
-i \text{Tr} \int_0^\infty dt_1 dt_2e^{-\epsilon t_1}\big((2-\eta_N)e^{-t_2^2-t_2^{-2}}+ 2 \; \frac{d}{dt_2}(e^{-t_2^2-t_2^{-2}})\big)\times\\ \times \frac{(-i)}{(4\pi)^2(t_2+t_2)^2}\left(1+\frac{i \bar R}{6}(t_1+t_2)\right)\Bigg|_{\epsilon\to -2 i \lambda_k}\,.
\end{aligned}
\ee

As far as the first order term (and for the second order as well) is concerned, both the minimal   and the non-minimal heat kernel expansions have to be applied. As an illustration, the minimal term up to first order reads
 \ba
&&-\int_0^\infty dt_1 dt_2 dt_3  e^{-\epsilon (t_1+t_3)}\big((2-\eta_N)e^{-t_2^2-t_2^{-2}}+ 2 \; \frac{d}{dt_2}(e^{-t_2^2-t_2^{-2}})\big)\cdot \nonumber\\
\label{3.14}
&& \qquad \qquad \cdot\;\;(\int_0^\infty dt_4 e^{iyt_4}e^{-t_4^2-t_4^{-2}}\frac{(-i)}{(4\pi)^2(t_1+t_2+t_3+t_4)^2}\left(1+\frac{i \bar R}{6}(t_1+t_2+t_3+t_4)\right) +\\
&&\qquad\qquad \qquad\qquad\qquad\qquad\qquad+ u_k\frac{(-i)}{(4\pi)^2(t_1+t_2+t_3)^2}\left(1+\frac{i \bar R}{6}(t_1+t_2+t_3)\right))\Bigg|_{\epsilon\to -2 i \lambda_k}\nonumber\;.
\ea
and analogously for the non-minimal term with \eqref{3.12}.
Reabsorbing the heat kernel coefficients $\Omega_i$ in the definition of the potential $u_k$, we report here the terms  we will be using up to first order in curvature expansion:
\ba\label{3.15}
&\text{Tr}(K^{-1}u_k e^{i \bar \Box t})^{\text{minimal}} &\approx\;\; \frac{(-i)}{(4\pi t)^2}\left(-\frac{4}{3}\Lambda_k - 3 \bar R - i\frac{2}{9}\Lambda_k \bar R t \right)\;,\\
\label{3.16}
&\text{Tr}(K^{-1}u_k e^{i \bar \Box t})^{\text{non-minimal}} &\approx\;\; \frac{(-i)}{(4\pi t)^2}\left(\frac{5}{t}+i \frac{5}{12}\bar R\right)\;,\\
\label{3.17}
&\text{Tr}((K^{-1}u_k)^2 e^{i \bar \Box t})^{\text{minimal}} &\approx\;\; \frac{(-i)}{(4\pi t)^2}\left(\frac{16}{9}\Lambda_k^2 + i \frac{8}{27}\bar R t \right)\;,
\\\label{3.18}
&\text{Tr}((K^{-1}u_k)^2 e^{i \bar \Box t})^{\text{non-minimal}}&\approx\;\; \frac{(-i)}{(4\pi t)^2}\left(-\frac{2}{3t}\Lambda_k - \frac{3}{2t} \bar R - i\frac{1}{18}\Lambda_k \bar R \right)\;.
\ea
where $\approx$ denotes that the right hand side is correct up to higher 
orders in curvature invariants.

\subsection{Evaluation of the heat kernel time integrals}
\label{s3.2}
In the previous section we have arrived at a closed convergent proper time 
expression for the functional trace of the r.h.s of flow equation 
\eqref{3.1}. At this stage, we have to solve the proper time integrals. 
We are instructed to compute them for $\epsilon>0$ and 
then to analytically continue to $-2i\lambda_k$. This would be easy 
if the integral would be analytically computable but it is not. We could 
numerically integrate it and then fit a function analytic in $\epsilon$ 
to approximate it. In this exploratory paper we will be content with 
the following very crude approximation which is the better the larger 
$\epsilon$ and which has the advantage to produce a closed expression:
We approximate $\forall n  \geq 0$ the following integrals as follows:
\ba\label{3.19}
\int_0^\infty dt_1 \int_0^\infty dt_2 e^{-\epsilon t_1} e^{-t_2-t_2^{-2}}\frac{1}{(t_1+t_2)^n} &\approx& \int_0^\infty dt_1 e^{-\epsilon t_1} \int_0^\infty dt_2 e^{-t_2-t_2^{-2}}\frac{1}{(t_2)^n}\\ \nonumber
&=& \frac{1}{\epsilon}\int_0^\infty dt_2 e^{-t_2-t_2^{-2}}\frac{1}{(t_2)^n}\,.
\ea
Using this approximation we are able to evaluate all the heat kernel 
integrals numerically and to perfor the 
analytic continuation explicitly. The singlarity of this approximation
as $\epsilon\to 0$ is incorrect for $n\ge 2$ thus the exact flow
will be somewhat better behaved at $\lambda_k\to 0$ than we can 
compute at the moment.
  
We will denote by $I_{m,n}$ ($J_{m,n}$) the integrals involving $m$ powers of the cut-off regulator functions (for $J_{m,n}$ the first function is derived wrt. the heat kernel time) and the $n$-th power of the proper time in the denominator:
\ba
I_{m,n}&=&\int_0^\infty dt_1\cdots dt_m \frac{e^{-t_1^2-t_1^{-2}}\cdots e^{-t_m^2-t_m^{-2}}}{(t_1+\cdots +t_m)^n}\;,\\
\label{3.22}
J_{m,n}&=&\int_0^\infty dt_1\cdots dt_m \frac{\frac{d}{dt_1}(e^{-t_1^2-t_1^{-2}})\cdots e^{-t_m^2-t_m^{-2}}}{(t_1+\cdots +t_m)^n}\;.
\ea
As an illustration, the integrals for $m=1$ can be solved analytically
\ba\label{3.21}
I_{1,n}&=&\int_0^\infty dt \frac{e^{-t^2-t^{-2}}}{t^n} = K_{\frac{n-1}{2}}(2)\;,\\
\label{3.22}
J_{1,n} &=&\int_0^\infty dt \frac{d}{dt}\frac{e^{-t^2-t^{-2}}}{t^n} =n K_{\frac{n}{2}}(2)\;.
\ea
where $K$ is the modified Bessel function. We will also list some useful numerical result
\ba\label{3.23}
 I_{2,2}&=&\int_0^\infty dt_1 dt_2 \frac{e^{-t_1^2-t_1^{-2}}e^{-t_2^2-t_2^{-2} }}{(t_1+t_2)^2} \approx 0.00308\;,\\
 \label{3.24}
J_{2,2}&=&\int_0^\infty dt_1dt_2 \frac{\frac{d}{dt_1}(e^{-t_1^2-t_1^{-2}})e^{-t_2^2-t_2^{-2}}}{(t_1+t_2)^2} \approx0.00310\;.
\ea

It is important to emphasize at this stage, that due to the approximation \eqref{3.19} in the evaluation of the integrals
the flow will contain additional terms of $\lambda_k$ in the denominator as in the standard FRG-ASQG computation. This will prevent us from
taking the vanishing $\lambda_k \to 0$ limit.

\subsection{Beta functions anf flow equations}
\label{s3.3}
Having evaluated the traces, we can now come back to \eqref{3.1} and 
compare with the l.h.s. of the equation \eqref{2.12}. In particular, 
having disentangled the flow of $\kappa_k$, we will be left with the 
flow of the two (dimensionless) gravitational coupling constants. 
Those can be identified by comparing on the l.h.s. and the r.h.s. 
the terms proportional to the identity operator 
(furnishing $k \partial_k \lambda_k$) and those proportional to the 
Ricci scalar (furnishing $k \partial_k g_k$). 
The flow of the two dimensionless coupling constants read:
\ba\label{3.25}
k \partial_k \lambda_k&=& -4\lambda_k + \eta_N \lambda_k - \frac{g_k}{4\pi}\frac{1}{2  \lambda_k} \left((2-\eta_N)\left(I_{1,2}+\frac{1}{2\lambda_k}\left(I_{2,2}-5I_{1,3}+\frac{4}{3}\lambda_kI_{1,2}\right) \right.\right.
\\
&&\left.\left.+ \frac{1}{(2\lambda_k)^2}\left(I_{3,2}+2\left(I_{2,2}-5I_{1,3}+\frac{4}{3}\lambda_kI_{1,2}\right)-i\frac{16}{9}\lambda_k^2I_{1,2}+ \frac{2}{3}I_{1,3} \right)\right) \right.\nonumber\\
&&
\left.+2\left(J_{1,2}+\frac{1}{2\lambda_k}\left(J_{2,2}-5J_{1,3}+\frac{4}{3}\lambda_kJ_{1,2}\right) \right.\right.\nonumber\\
&&
\left.\left.\qquad+ \frac{1}{(2\lambda_k)^2}\left(J_{3,2}+2\left(I_{2,2}-5J_{1,3}+\frac{4}{3}\lambda_kJ_{1,2}\right)-i\frac{16}{9}\lambda_k^2J_{1,2}+ \frac{2}{3}J_{1,3} \right)\right)\right)\nonumber\;,
\ea

\ba\label{3.26}
k \partial_k g_k &=& 2g_k - \frac{g_k^2}{2\pi}\frac{1}{2  \lambda_k} \left((2-\eta_N)\left(\frac{i}{6}I_{1,1}+\frac{1}{2  \lambda_k}\left(\frac{i}{6}I_{2,1}+3I_{1,2}+\frac{2i}{9}\lambda_ kI_{1,1}-\frac{5i}{12}I_{1,2}\right) \right.\right.
\\\nonumber
&&\left.\left.+ \frac{1}{(2  \lambda_k)^2}\left(\frac{i}{6}I_{3,1}+2\left(\frac{i}{6}I_{2,1}+3I_{2,2}+i\frac{2}{9}\lambda_ kI_{2,1}-i\frac{5}{12}I_{2,2}\right)\right.\right.\right.\\\nonumber
&&\left.\left.\left.\qquad-i\frac{8}{27}I_{1,1}+ \frac{3}{2}I_{1,3}+ i \frac{1}{18}\lambda_kI_{1,2} \right)\right) \right.\\\nonumber
&&
\left.+2\left(\frac{i}{6}J_{1,1}+\frac{1}{2  \lambda_k}\left(\frac{i}{6}J_{2,1}+3J_{1,2}+i\frac{2}{9}\lambda_ kJ_{1,1}-i\frac{5}{12}J_{1,2}\right)\right.\right.\\\nonumber
&&
\left.\left.\qquad+ \frac{1}{(2  \lambda_k)^2}\left(\frac{i}{6}J_{3,1}+ 2\left(\frac{i}{6}J_{2,1}+3J_{2,2}+i\frac{2}{9}\lambda_ kI_{2,1}-i\frac{5}{12}J_{2,2}\right)\right.\right.\right.\\ \nonumber &&\left.\left.\left.\qquad-i\frac{8}{27}J_{1,1}+ \frac{3}{2}J_{1,3}+ i \frac{1}{18}\lambda_kJ_{1,2}\right)\right)\right)\;.
\ea
Recalling that $\eta_N = k\partial_kg_k/g_k$ we find the explicit expression for the beta functions. These are polynomials in $g_k$ and $\lambda_k$ and have structurally the following form
\ba\label{3.27}
\beta_\lambda &=& \frac{a_1\lambda_k^7+g_k(a_2 \lambda_k^3+ a_3 \lambda_k^4+a_4 \lambda_k^5) + g_k^2(a_5+a_6\lambda_k+a_7\lambda_k^2+a_8\lambda_k^3+a_9\lambda_k^4)}{\lambda_k^6+g_k (a_{10}\lambda_k^3+a_{11} \lambda_k^4+a_{12} \lambda_k^5)}\;,\\\label{3.28}
\beta_g &=& \frac{ c_1 g_k \lambda_k^3+ g_k^2 (c_2 + c_3 \lambda_k+ c_4 \lambda_k^2)}{g_k(c_5 + c_6 \lambda_k + c_7 \lambda_k^2)+ \lambda_k^3 }\;.
\ea
where the $a$'s and the $c$'s are complex numerical coefficients. We note that the small$-g_k$ expansion presents the behaviour
\ba\label{3.29}
\beta_\lambda &=& -2 \lambda_k- \frac{g_k}{\lambda_k^3}\text{Pol}[ 1,\lambda_k,\lambda_k^2] + O(g_k^3)\;,\\
\label{3.30}
\beta_g &=& 2 g_k - \frac{g_k^2}{\lambda_k^3}\text{Pol}[ 1,\lambda_k, \lambda_k^2, \lambda_k^3]+ O(g_k^3)\;,
\ea
which corresponds to the expected near-perturbative regime, except for the singularity at $\lambda_k \to 0$.

\subsection{UV Fixed points of the dimension free flow}
\label{s3.4}
Looking for the fixed points,  the two beta function \eqref{3.25} and \eqref{3.26} vanish when we set $g_* = 0$ and afterwards take the limit $\lambda_*\to0$. However, if one sets before $\lambda_*=0$, then they diverge with an inverse power of $\lambda_k$. This is the cost to pay for our approximation in solving the proper time integral.

Furthermore, one can find that they vanish also when $k \to \infty$ for
\be\label{3.33}
\lambda_* = 0.460 +  0.050 \;i\;, \qquad g_* =  1.013 + 0.420\;i  \,,
\ee
reaching the analogue of the Reuter fixed point \cite{Martin} in Lorentzian spacetimes. Furthermore, we observe also that the  anomalous dimension of the dimensionful Newton's constant  is $1.975+ 4.756 i$ at the UV fixed point, whose real part is very close to the value of 2 found in Euclidian ASQG.

The set of the two complex valued 
beta function can be rewritten as a set of four real valued 
beta functions, by decomposing $\lambda_k$ and $g_k$ 
into their real and imaginary parts and also decomposing the original beta 
functions into their real and imaginary contributions. This yields 
four real valued flow equations for four real valued parameters. 
In order to understand the nature of the fixed point, we can
pick two out four paramewters at their fixed point values shown in 
\eqref{3.33} and plot
the flow of the remaining ones for initial data in the vicinity of their 
fixed point values at $k=k_0=\bar{k}=1$
(see Figures \ref{fig:1}, \ref{fig:2}, \ref{fig:3}). Note that this does not 
yet numerically prove attractive nature of the fixed point for 
initial data in a full four dimensional neighbourhood, we investigate 
this more complicated problem in the next subsection. 
\begin{figure}[]
	\centering
	\includegraphics[width=.47\textwidth]{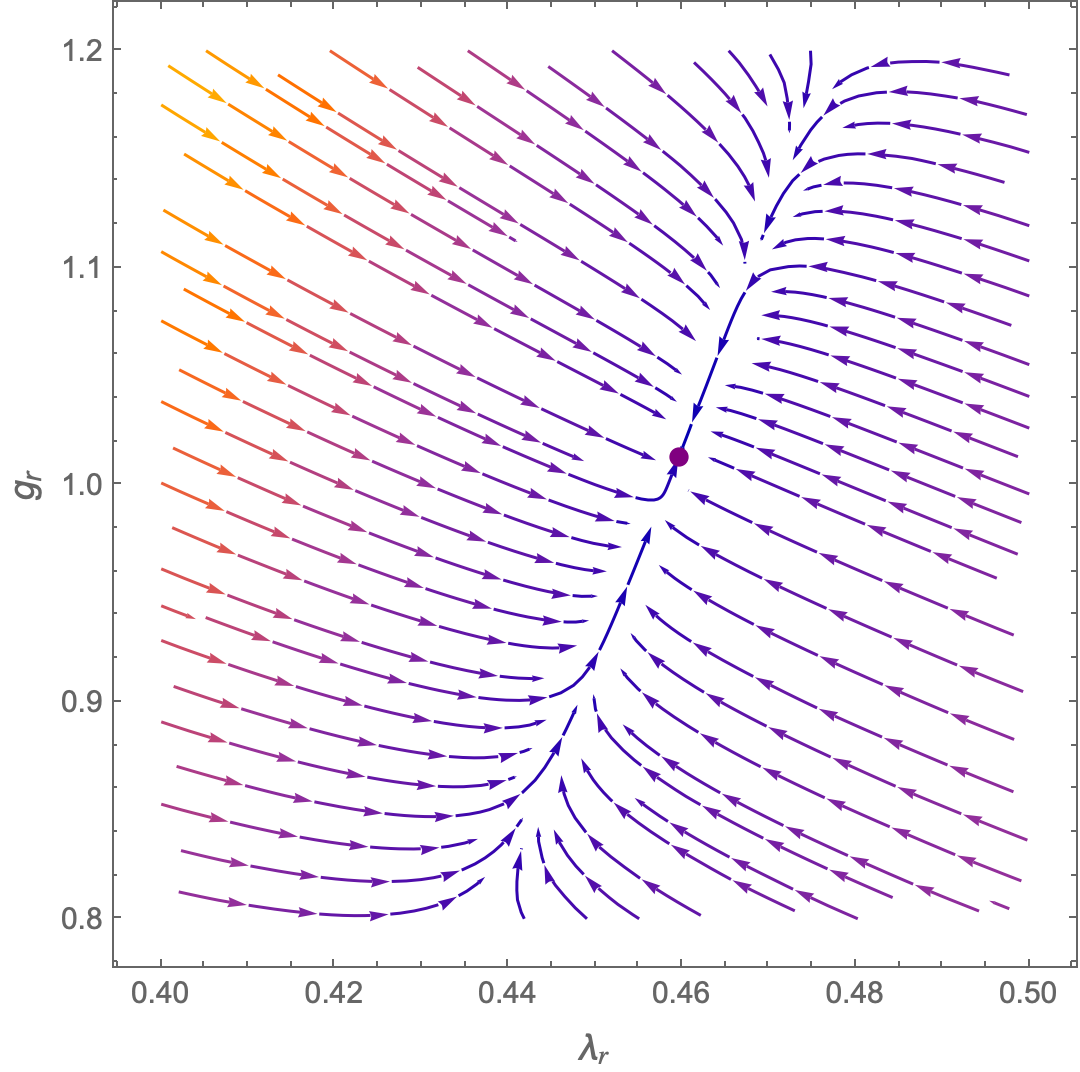}\qquad	\includegraphics[width=.47\textwidth]{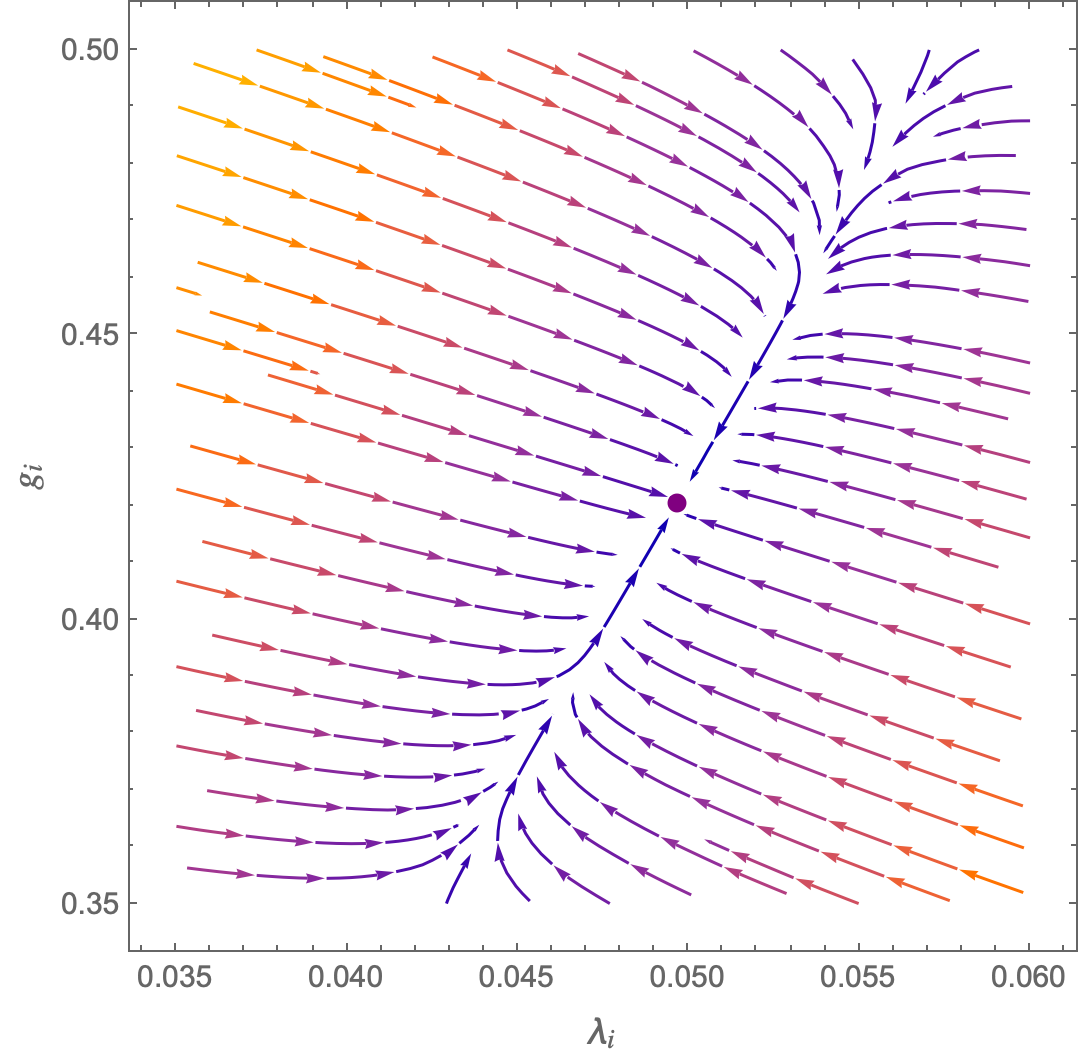}
	\caption{Projected flow diagram of the real (left) and imaginary (right) part in the $\lambda-g$ plane. The purple dot represents the fixed point in \eqref{3.33}. The arrows point towards an increasing $k$, hence at $k \to \infty$ the fixed point is attractive in both projections.}
	\label{fig:1}
\end{figure}
\begin{figure}[]
	\centering
	\includegraphics[width=.47\textwidth]{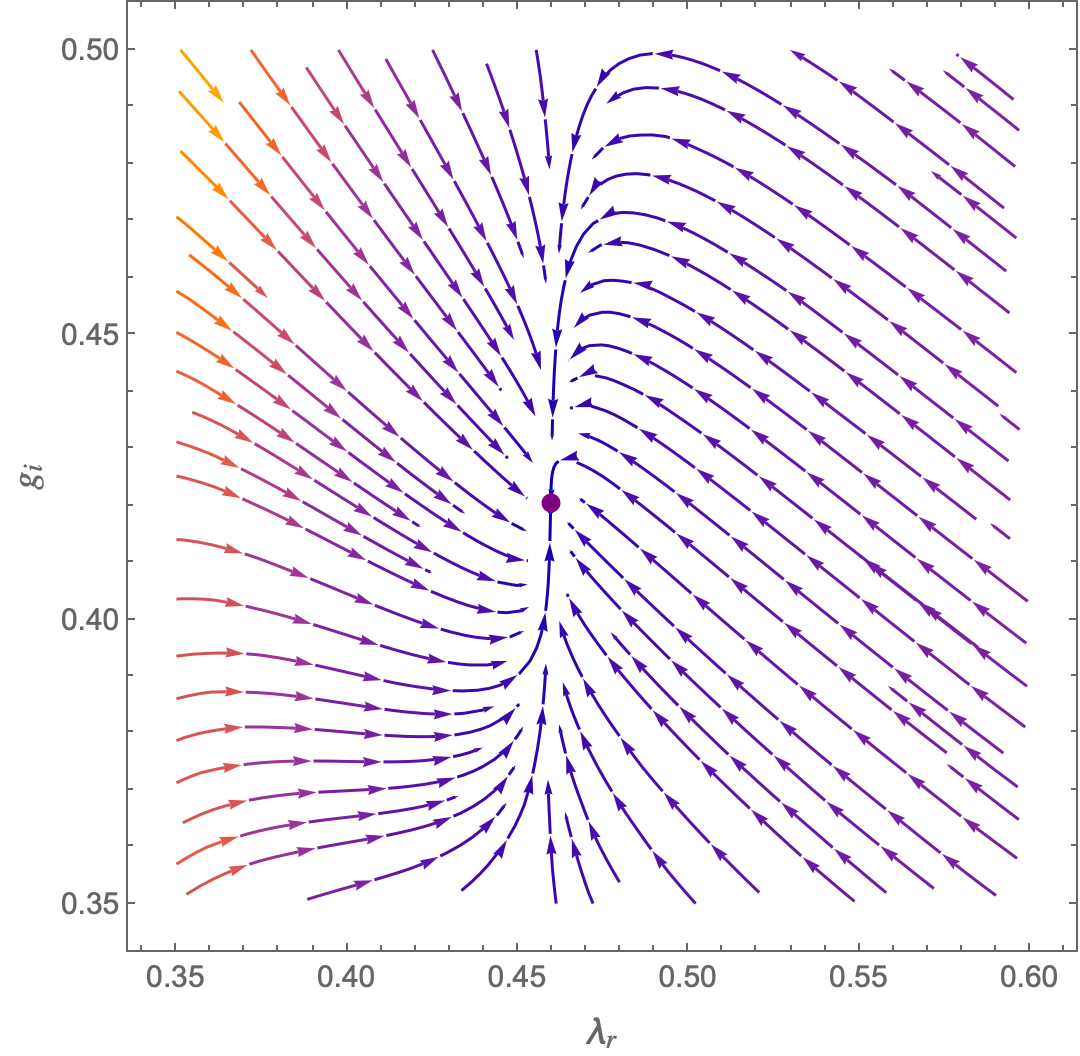}\qquad	\includegraphics[width=.47\textwidth]{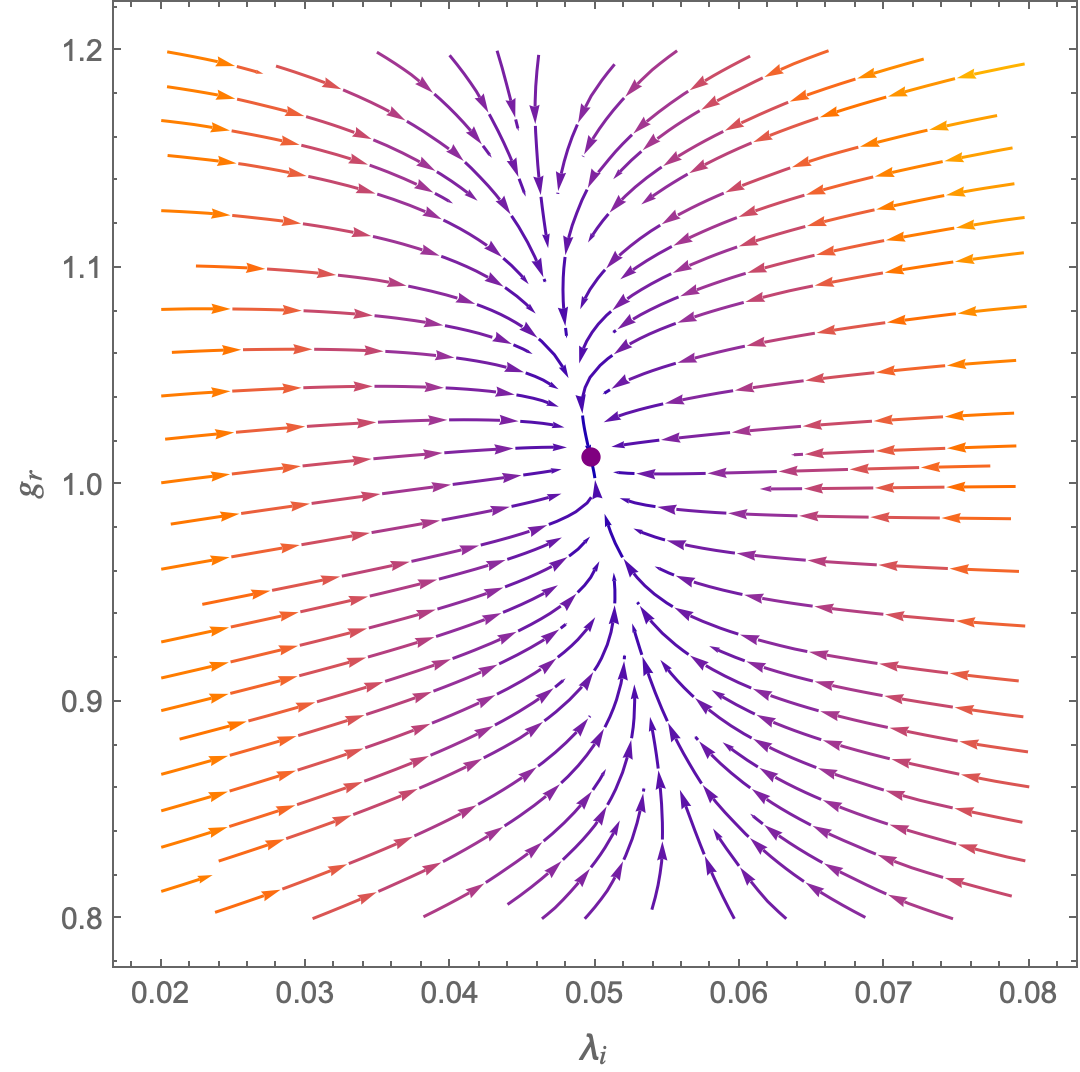}
	\caption{Flow diagram of the $\lambda_\text{real}- g_\text{imaginary}$ (left) and $\lambda_\text{imaginary}- g_\text{real}$ (right) part. The arrows along the trajectories point towards increasing value of $k$, and that means that the trajectories flow into the fixed point   \eqref{3.33} in the UV.}
	\label{fig:2}
\end{figure}
\begin{figure}[]
	\centering
	\includegraphics[width=.47\textwidth]{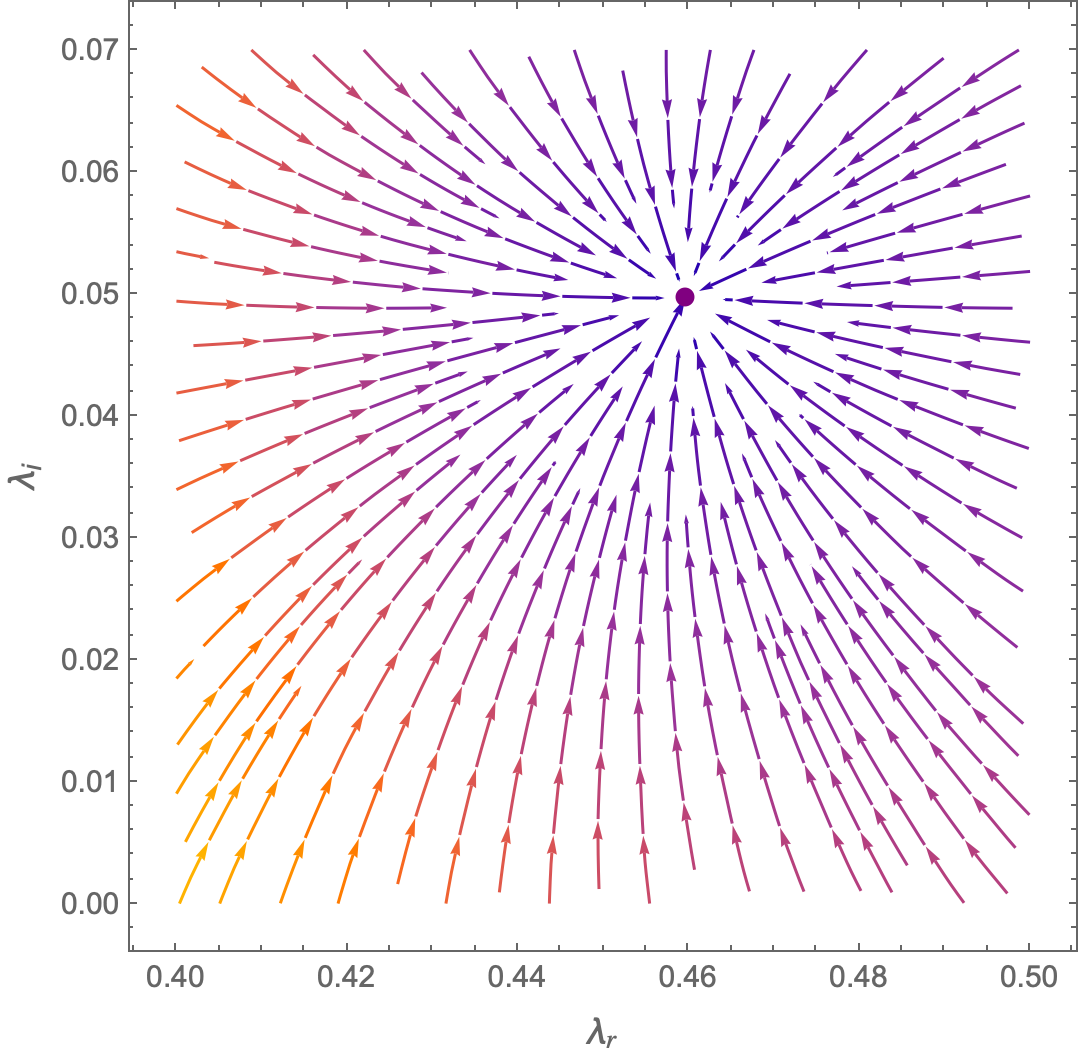}\qquad	\includegraphics[width=.47\textwidth]{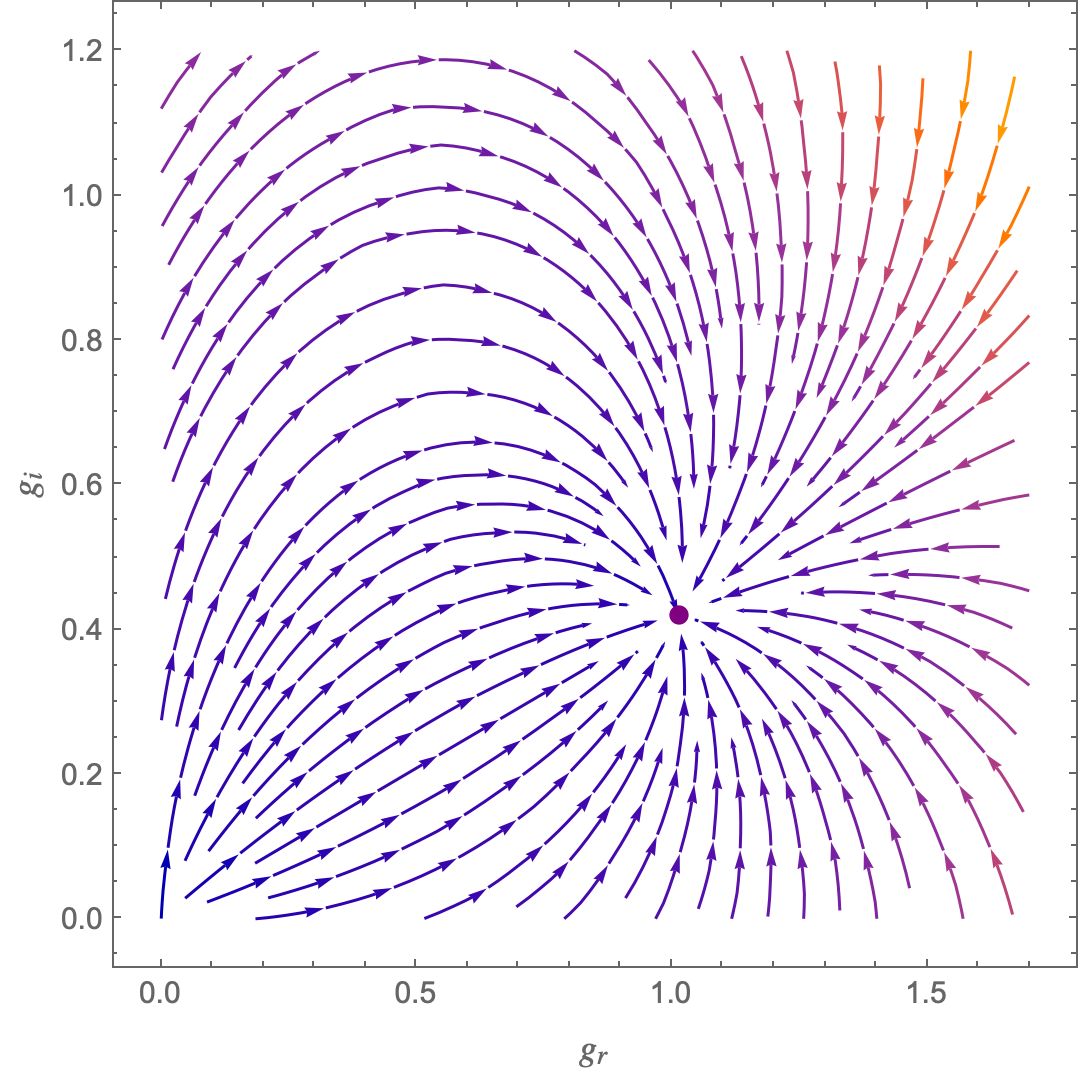}
	\caption{Flow diagram in the $\lambda_\text{real}- \lambda_\text{imaginary}$ (left, projected to $g = g_\ast$) and in the $g_\text{real}- g_\text{imaginary}$ (right, projected to $\lambda = \lambda_\ast$) space. The purple dot  depicted is the fixed point in \eqref{3.33}. The arrows point towards $k \to \infty$.}
	\label{fig:3}
\end{figure}
In all 6 flow projections studied, the trajectories flow into the fixed 
point for $k \to \infty$.

Another interesting quantity carrying information about the nature of the 
fixed point are the critical exponents. 
The critical exponents can be computed by linearizing the flow around the 
fixed point, computing the stability matrix and determining its 
eigenvalues. The critical exponents for the Lorentzian UV fixed point 
\eqref{3.33} are
\be\label{3.34}
\theta_1 = 12.24 - 0.07 \;i \;, \qquad \theta_2 = 0.95 + 0.017 \;i\,,
\ee
therefore, being their real part positive, both coupling constants  are associated to two relevant 
directions. Here the convenvention is that in the diagonalised form 
the couplings behave near the fixed point as $g_j(k)-g_j^\ast\propto 
(k_0/k)^{\theta_j}$ where $k_0$ is the point at which one sets 
initial conditions. That is, for $\Re(\theta_j)>0$ the fixed point is 
reached insensitive to the initial condition, the coupling must not be 
fine tuned and thus must be measured (therefore it is relevant).   
Note that we do not expect here the critical exponents to be complex conjugated as in the standard FRG-ASQG treatment because of the intrinsically complex nature of the flow.

\subsection{IR limit of the dimensionful couplings and admissible trajectories}
\label{s3.5}
Being interested in the full effective action, 
which corresponds to the $k \to 0$ limit, it is important to 
prove the existence of admissible trajectories, 
i.e., those trajectories for which $G_{k=0}=\text{real}$ and 
$\Lambda_{k=0}=\text{real}$. 
Fixing an arbitrary initial condition at a chosen scale $\bar k = 1$ 
for $\lambda_\text{real}$ and $g_\text{real}$ 
(preferably closed to the UV fixed point) we can integrate down the flow 
and fine tune the initial data value of $\lambda_\text{imaginary}$ and 
$g_\text{imaginary}$ s.t. $\text{Im}[G_{k = 0}] = 0$ and 
$\text{Im}[\Lambda_{k = 0}] = 0$. This fine tuning is equivalent 
to computing the maps 
\ba
\lambda_\text{imaginary}\big|_{k = \bar k} &= &f_1(\lambda_\text{real}, g_\text{real} )\big|_{k = \bar k} \;,\label{3.36}\\
g_\text{imaginary}\big|_{k = \bar k} &=& f_2(\lambda_\text{real}, g_\text{real} )\big|_{k = \bar k}\;,\label{3.37}
\ea
thereby reducing the flow by a dimension of 2. We have just started to 
investigate this very interesting question which has to be performed 
numerically. A priori, it could be that there are domains 
in the real $(g,\lambda)$ plane such that there exists precisely 
one, several or no solution of the fine tuning problem in the imaginary 
$(g,\lambda)$ plane.\\ 
\\
In what follows we summarise our present numerical findings: \\
\\
First, we prove numerically the existence of 
admissible trajectories for selected initial conditions.
\begin{figure}[]
	\centering
	\includegraphics[width=.9\textwidth]{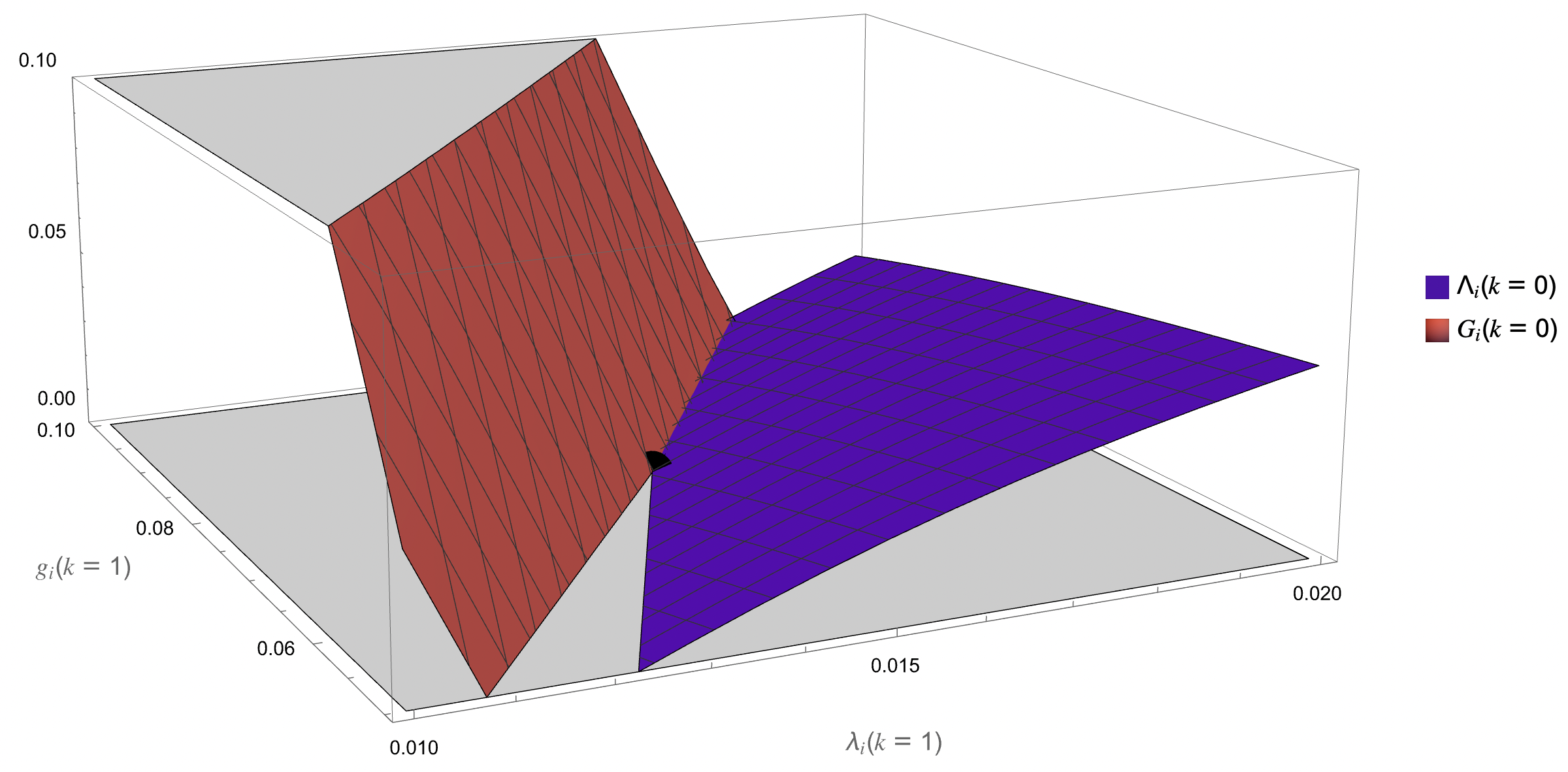}
	\caption{Graphic method for the proof of the existence of an admissible trajectory for a given set of real initial conditions. At fixed $\bar k =1$ we choose the fixed point set  of initial conditions $g_r = 1.013$ and $\lambda_r = 0.460$, we integrate down the flow to $k=0$ and we plot the surface of the dimensionful imaginary part of $G$ and $\Lambda$ when $k \to 0$. The two surfaces intersect exactly in one point on the plane $\text{Im}[\Lambda_{k = 0}] = \text{Im}[G_{k = 0}] = 0$:  the intersection point furnishes the corresponding pair of imaginary initial conditions at $\bar k = 1$ for $g_i$ and $\lambda_i$ giving rise to an admissible trajectory.}
	\label{fig:8}
\end{figure}
As an example, we select a  trajectory with initial conditions 
$\lambda_\text{real}(k = 1) \approx \lambda_\ast$  and 
$g_\text{real}(k = 1) \approx g_\ast$
at $\bar{k}=1$.
By means of a graphic method illustrated in Fig. \ref{fig:9} we find the 
corresponding $\lambda_\text{imaginary}(k = 1) = 0.015$ and 
$g_\text{imaginary}(k = 1) =  0.078$, realizing an admissible trajectory. 
In Figures \ref{fig:4} and \ref{fig:5} the flow of the dimensionful 
couplings of this selected admissible trajectory is depicted.
Furthermore, we tested that both real and imaginary parts of this 
trajectory flow into the UV fixed point. This can be appreciated in the 
plots of the dimensionless coupling in Figures \ref{fig:6} and \ref{fig:7}.
\begin{figure}[]
	\centering
	\includegraphics[width=.47\textwidth]{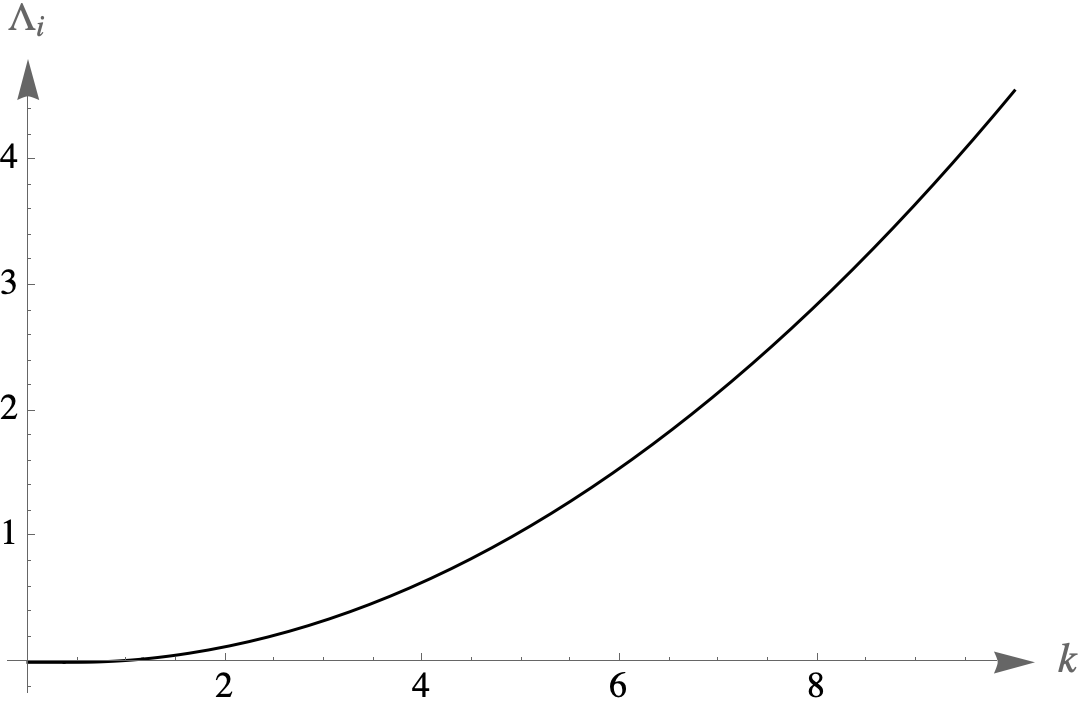}\qquad	\includegraphics[width=.47\textwidth]{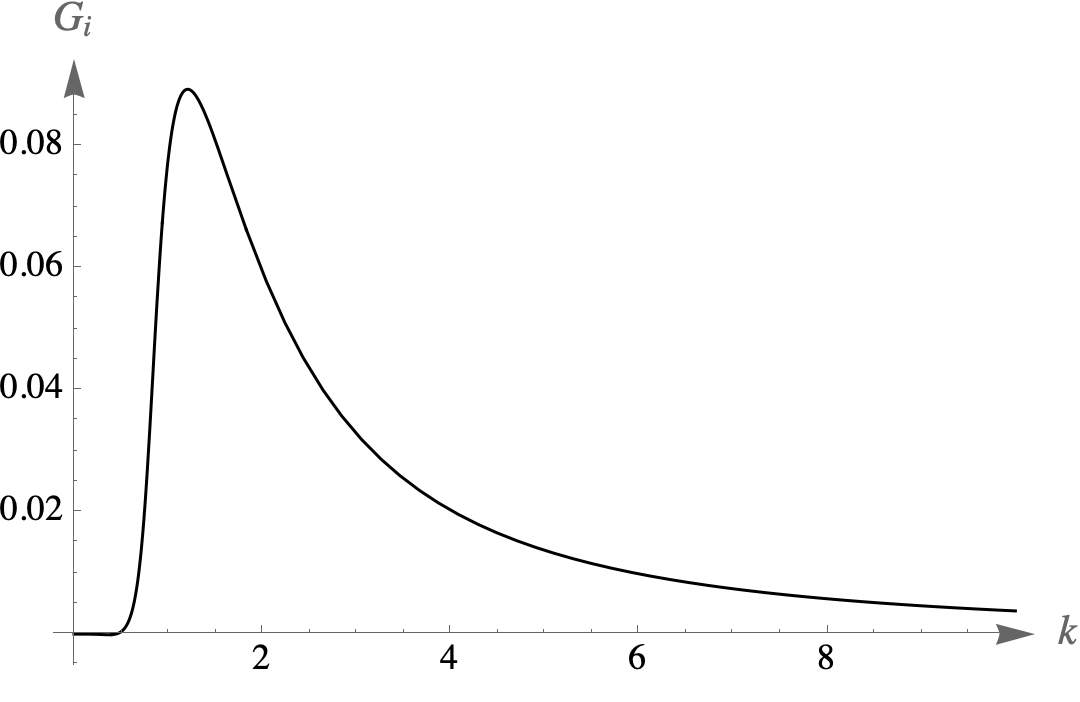}
	\caption{Admissible trajectory with 
        $\lambda(k = 1) = 0.460+0.015i$  
        and $g(k = 1) = 1.013+0.079i$. The imaginary parts of 
        the dimensionful coupling constants are vanishing for $k \to 0$.}
	\label{fig:4}
\end{figure}
\begin{figure}[]
	\centering
	\includegraphics[width=.47\textwidth]{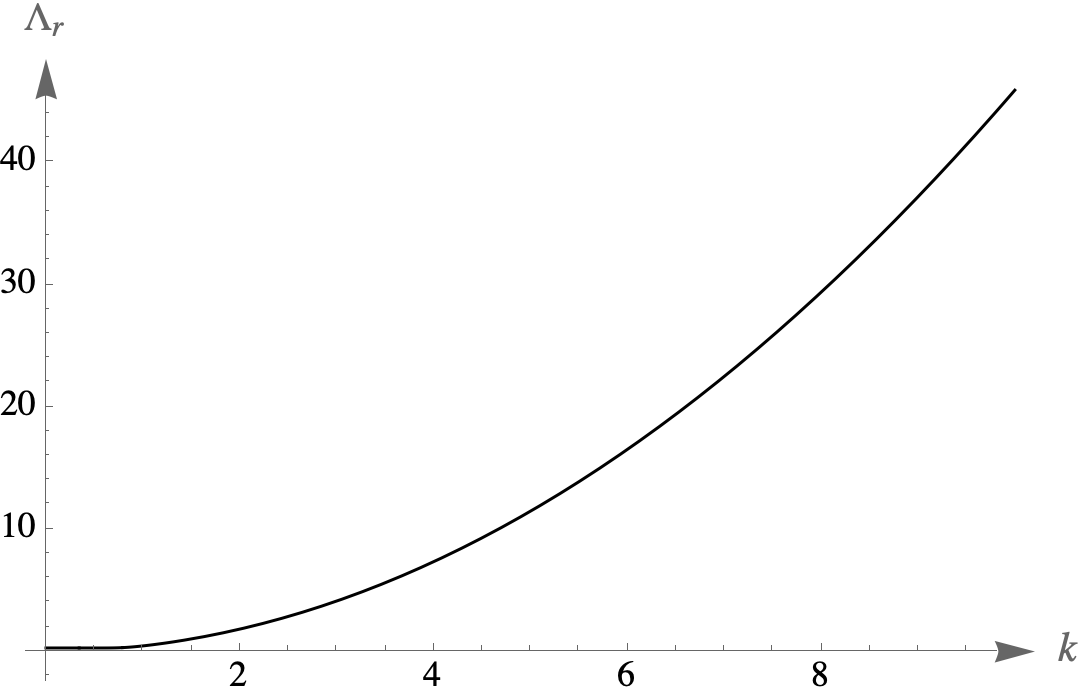}\qquad	\includegraphics[width=.47\textwidth]{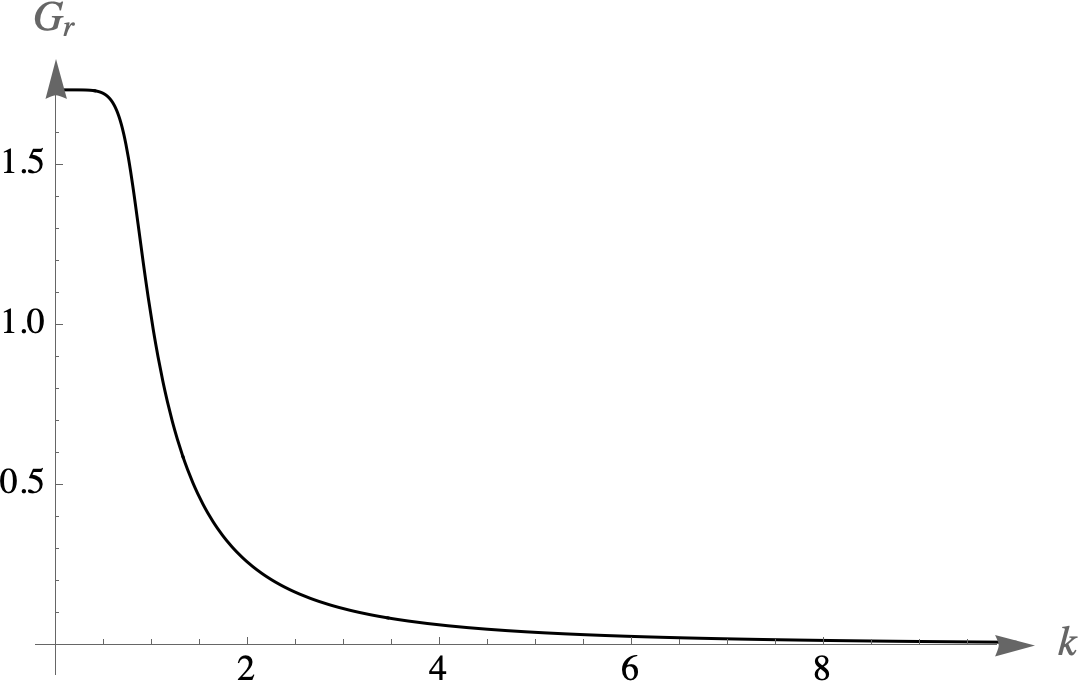}
	\caption{Admissible trajectory with   $\lambda(k = 1) = 0.460+0.015i$  
		and $g(k = 1) = 1.013+0.079i$. The real part of the dimensionful coupling constants is well behaved and reaches a finite value (vanishes for $\Lambda_\text{real}$) when $k \to 0$.}
	\label{fig:5}
\end{figure}
\begin{figure}[]
	\centering
	\includegraphics[width=.47\textwidth]{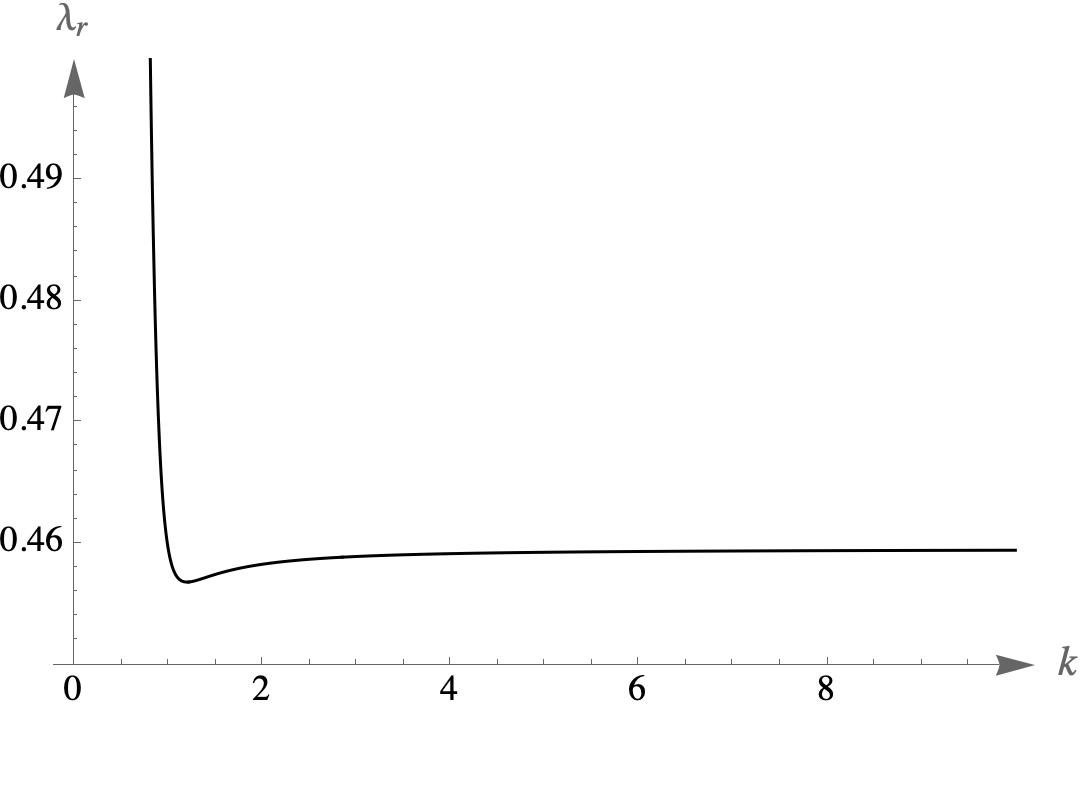}\qquad	\includegraphics[width=.47\textwidth]{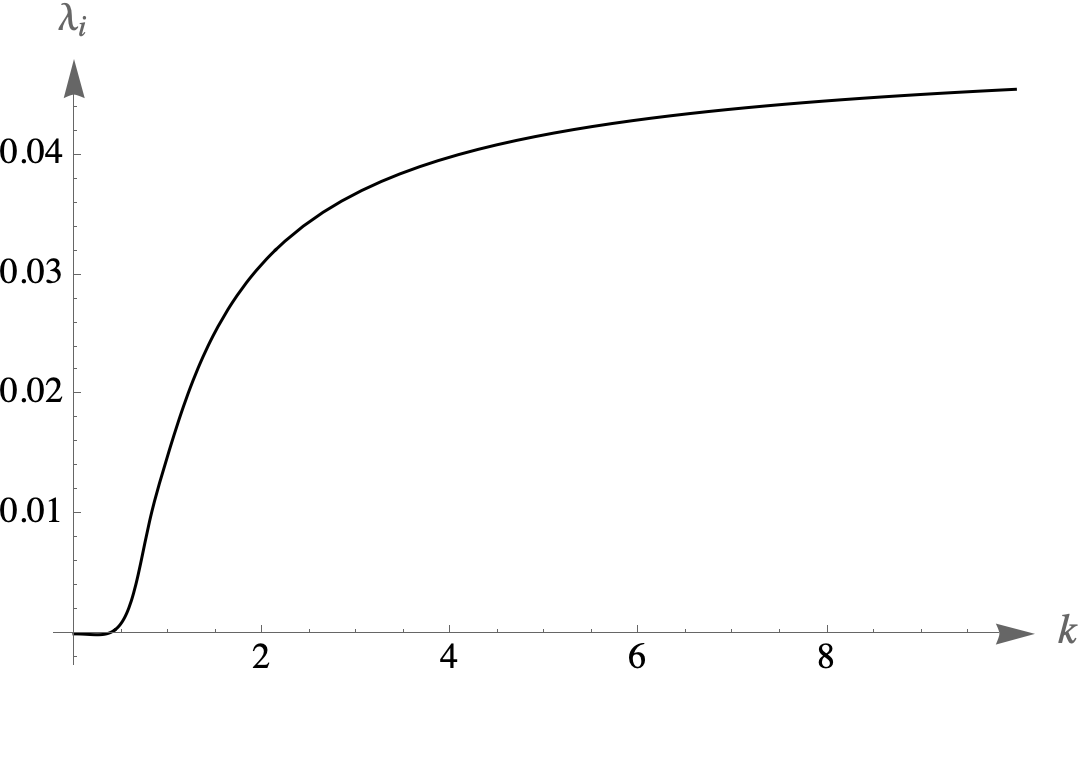}
	\caption{Admissible trajectory with   $\lambda(k = 1) = 0.460+0.015i$  
		and $g(k = 1) = 1.013+0.079i$. The flow of the real and imaginary parts of the dimensionless coupling $\lambda_k$ is reaching the UV fixed point when $k \to \infty$. Note the divergence for vanishing $k$ due to the approximation \eqref{3.19} performed in the evaluation of the integrals. }
	\label{fig:6}
\end{figure}
\begin{figure}[]
	\centering
	\includegraphics[width=.47\textwidth]{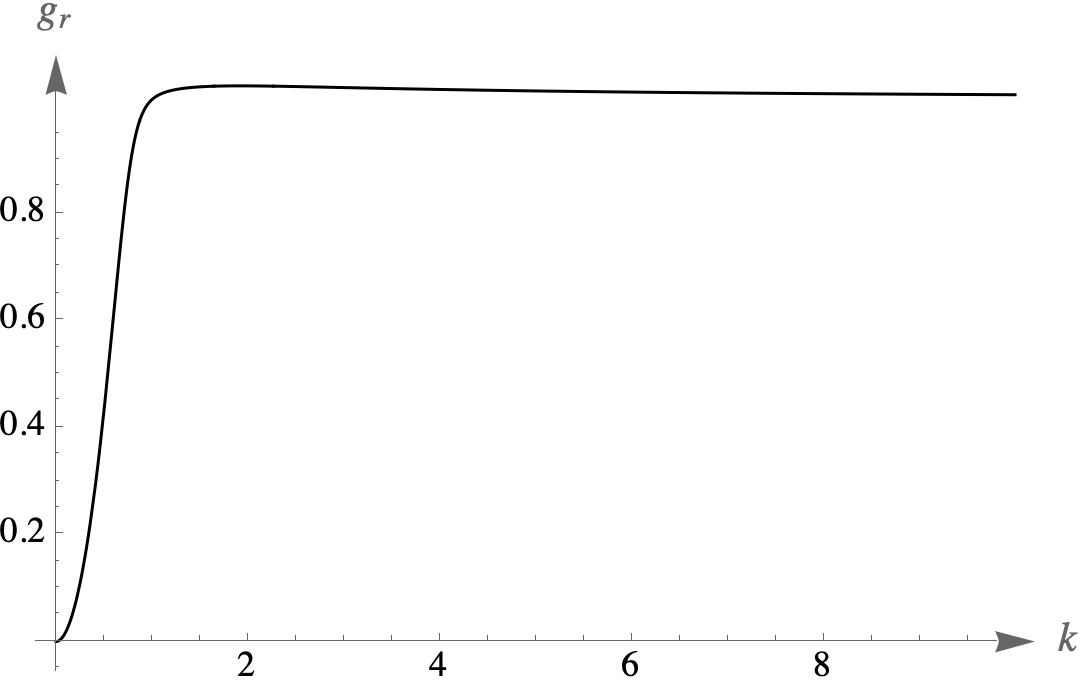}\qquad	\includegraphics[width=.47\textwidth]{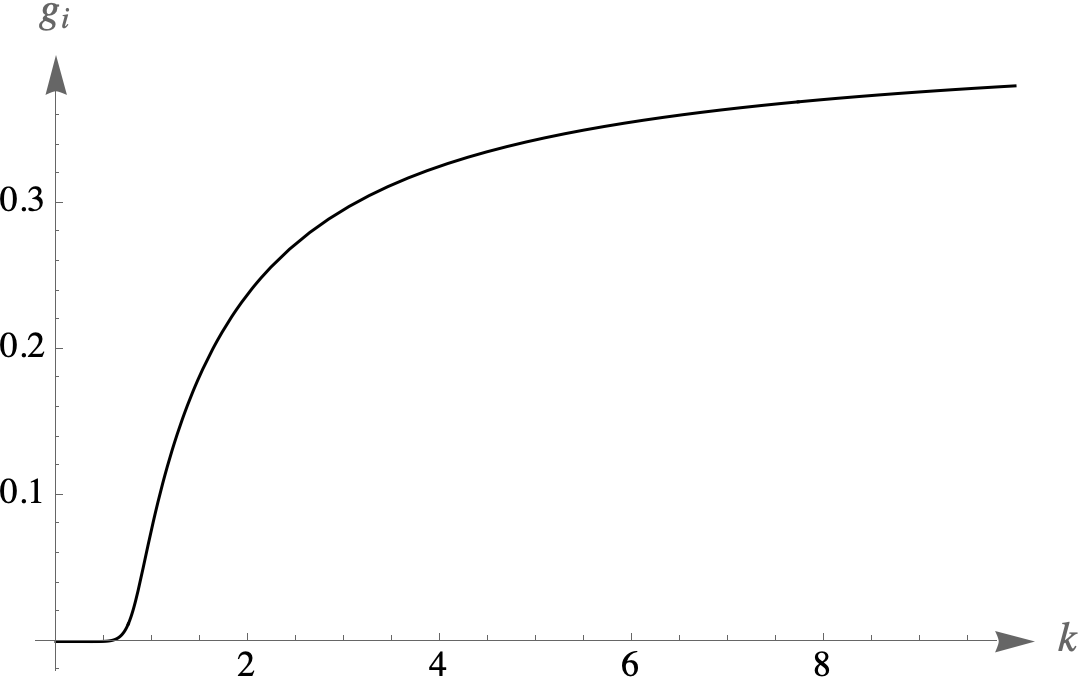}
	\caption{Admissible trajectory with   $\lambda(k = 1) = 0.460+0.015i$  
		and $g(k = 1) = 1.013+0.079i$. The real and imaginary parts of the dimensionless coupling $g_k$ flow into the UV fixed point when $k \to \infty$. In the IR both parts vanish.}
	\label{fig:7}
\end{figure}

Next, we
study the unicity and the existence of the solution to the system composed by 
\eqref{3.36} and \eqref{3.37} in the proximity of the fixed point 
(Fig. \ref{fig:9}).  We test the unicity of admissible trajectories for the 
region of initial parameters where they exist. Furthermore we observe the 
non-existence of trajectories for small values of $\lambda_r$: 
the existence of admissible trajectories seems to be related by a 
linear relation between $g_r$ and $\lambda_r$. However, this relation is 
non-trivial to find, because it is represents a relation between values 
of dimensionless and dimensionful couplings at different values of $k$.
Following the physical principle of the existence admissible trajectories, 
one should discard those region of parameter space where they do not exist. 
This allows to restrict the parameter space of initial conditions. 

Finally, as a side result, 
we simultaneously verified that the fixed point is attractive
in a full real four dimensional neighbourhood of initial data: Namely, 
it was necessary to 1. compute the flow for $k\in[1,\infty]$ for the four
dimesion free parameters $g_r,g_i,\lambda_r,\lambda_i$ in a four 
dimensional neighbourhood 
of their fixed point values, 2. to check that in each case 
this flow ends in the fixed 
point, 3. to compute the corresponding flow in $[0,1]$ for the dimensionful   
parameters $G_r,G_i,\Lambda_r,\Lambda_i$ (they have the same initial values as
$g_r,g_i,\lambda_r,\lambda_i$ at $k=1$) and 4. to determine for each initial
data pair $(g_r,\lambda_r)$ those initial data $(g_i,\lambda_i)$ for which 
$G_i=\Lambda_i=0$ at $k=0$ is reached, thereby constructing numerically the 
functions \eqref{3.36} and \eqref{3.37}.

\begin{figure}[]
	\centering
	\includegraphics[width=.9\textwidth]{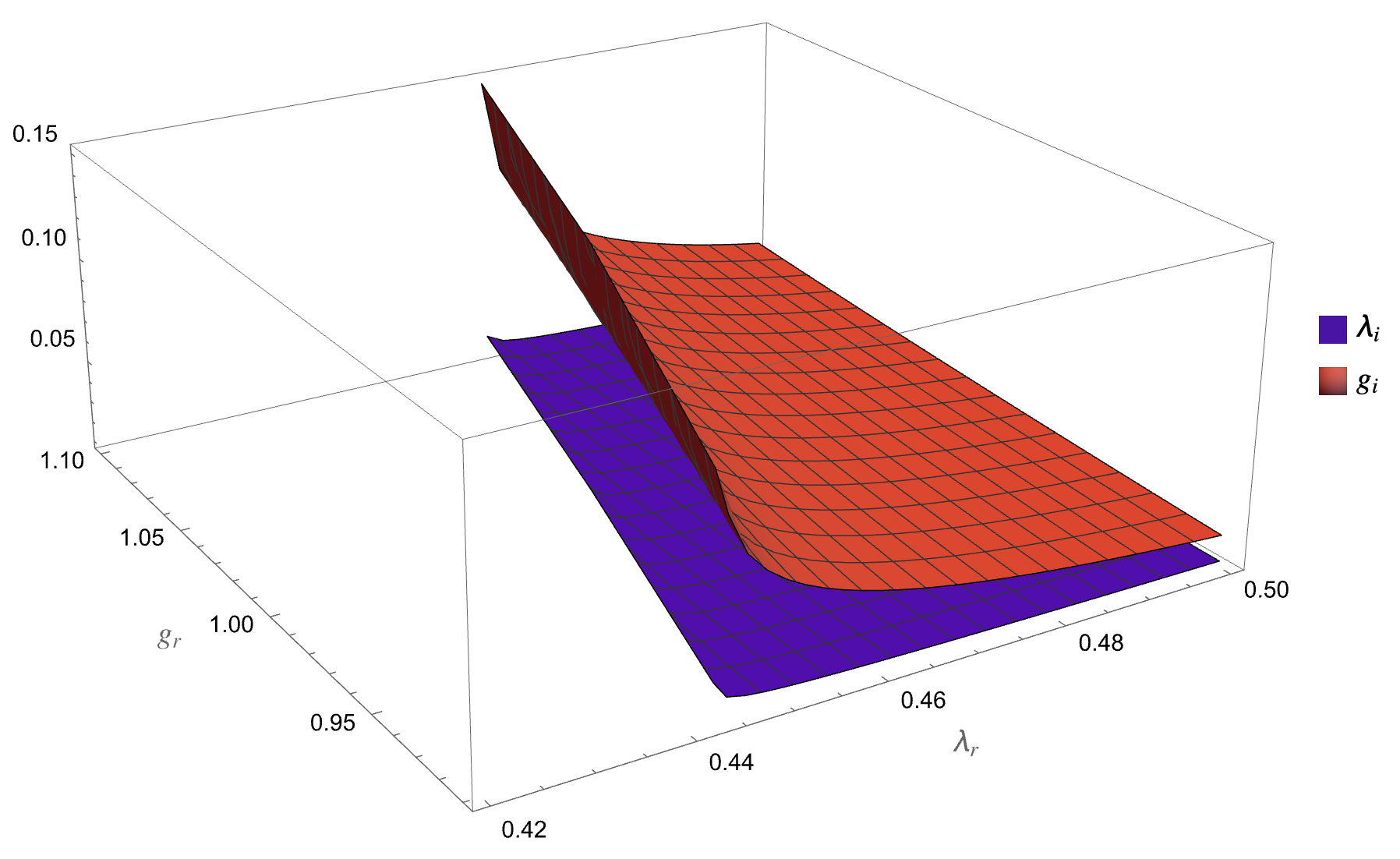}
	\caption{Plot of the functions $f_1$ and $f_2$ defined in \eqref{3.36} and \eqref{3.37} (here we fixed $\bar k = 1$). Points on the two surfaces realize admissible trajectories. Those exist for increasing values of $\lambda_r$ and cease to exist in the regime where no surface is depicted. Furthermore we see that the surfaces are regular surfaces, hinting at unicity of the admissible trajectories at fixed initial conditions.}
	\label{fig:9}
\end{figure}

\section{Conclusion}
\label{s4}

In this contribution we considered a certain Einstein-Klein-Gordon
theory as a showcase model to demonstrate that the ASQG and CQG approaches
can be fruitfully combined. In particular, CQG gives important input for how 
to actually define the class of EEA to start with, displaying new 
contributions coming from 1. the state underlying the Hamiltonian 
quantum theory, 2. 
measure factors coming from the momentum integrals, 3. 
restrictions on correlation functions of the true degrees of freedom 
only and 4. that Lorentzian signature is the most natural choice.

ASQG on the other hand offers a systematic
procedure for how to obtain a 
well defined effective action from which all the time ordered correlators 
of the Hamiltonian theory can be computed. The effective action
can be argued to be a 
complete definition of the theory. \\
\\
By exploiting the techniques for the Lorentzian heat kernel and 
introducing a new cut-off function, we computed the Lorentzian flow of an 
Einstein-Klein-Gordon model. Our analysis can be summarised in the 
following results:\\
1. We found that the coupling constant related to the matter 
contribution does not flow and also does not affect the flow of the 
gravitational coupling in the truncation considered here.
2. We computed the flow of Newton's constant and the cosmological constant 
and we found an attractive UV fixed point at the value 
$\lambda_* = 0.460 +  0.050 \;i\;,  g_* =  1.013 + 0.420\;i $. 
Furthermore, we computed the critical exponents and related the coupling 
constant to two relevant directions.\\
3. We proved the existence of admissible trajectories, 
integrating down the flow to $k \to 0$ and finding trajectories which 
flow from real valued dimensional couplings in the IR and reach the UV 
fixed point of the dimensionless couplings when $k \to \infty$.\\
\\
Among the many directions for future work are: Investigation of the 
space of admissible trajectories for the present model,
classification of the 
Lorentzian cut-off functions with respect to their necessary physical 
and mathematical properties, incorporation of the flow of the 
ghost matrix term (equivalently, one can avoid the gost matrix by 
using field redefinitions) and its dependence on the state 
$\omega$, higher order truncations, more realistic 
matter coupling and the Euclidian version which is in principle possible 
but more complicated except for very special matter such as \cite{12}.


\end{document}